\documentclass[%
 reprint,
 amsmath,amssymb,
 aps,
rmp,
]{revtex4-1}

\usepackage{graphicx}
\usepackage{dcolumn}
\usepackage{bm}
\usepackage{url} 

\begin{document}

\preprint{APS/123-QED}

\title{Experimental soft-matter science}

\author{Sidney R. Nagel}
\affiliation{%
 The James Franck and Enrico Fermi Institutes and the Department of Physics\\
The University of Chicago, Chicago, IL 60637}%




\date{\today}

\begin{abstract}
Soft materials consist of basic units that are significantly larger than an atom but much smaller than the overall dimensions of the sample.  The label ``soft condensed matter'' emphasizes that the large basic building blocks of these materials produce low elastic moduli that govern a material's ability to withstand deformations.  Aside from softness, there are many other properties that are also caused by the large size of the constituent building blocks.  Soft matter is  dissipative, disordered, far-from-equilibrium, non-linear, thermal and entropic, slow, observable, gravity-affected, patterned, non-local, interfacially elastic, memory-forming and active.  This is only a partial list of how matter created from large component particles is distinct from ``hard matter'' composed of constituents at an atomic scale. 

Issues inherent in soft matter raise problems that are broadly important in diverse areas of science and require multiple modes of attack.  For example, far-from-equilibrium behavior is confronted in biology, chemistry, geophysics, astrophysics and nuclear physics.  Similarly, issues dealing with disorder appear broadly throughout many branches of inquiry wherever rugged landscapes are invoked. 

This article reviews the discussions that occurred during a workshop held on January 30-31, 2016 in which opportunities in soft-matter experiment were surveyed.  Soft matter has had an exciting history of discovery and continues to be a fertile ground for future research.   
\begin{description}
\item[PACS numbers]
47.57.-s, 47.54.-r, 82.35.-x, 05.70.Ln, 46.65.+g, 83., 47.
\end{description}
\end{abstract}

\pacs{Valid PACS appear here}
\maketitle

\tableofcontents

\section{\label{sec:level1}Introduction:\protect\\}

\subsection{\label{sec:level2}Characteristics of soft condensed matter}

The label ``soft condensed matter'' does not do justice to the broad scope of this branch of research.  ``Softness'' emphasizes the low elastic moduli that govern a material's ability to withstand deformations.  However, that is only one attribute and many other properties go hand-in-hand with the malleability of a material.  

To understand some of the issues, it is relevant to be reminded of the common cause of the emergence of low elastic moduli in these soft materials.  In colloids, polymers, foams, granular matter, emulsions and other soft matter, the atoms are organized on a mesoscopic scale into entities that are much larger than an atom but still much smaller than the overall size of the material.  These large entities interact with one another to produce the elastic response of the material.  However, because the individual constituents are large, the moduli must necessarily be small.  This is essentially a dimensional-analysis argument \cite{Frenkel2002}.  The bulk and shear moduli of a material have the units [energy]/[volume].  While the interaction energy between two constituent particles is perhaps not exceedingly different from that between individual atoms, the volume of each particle is large.  Thus the moduli are orders of magnitude smaller than they would be for an atomic or molecular crystal. 

Thus materials made up of such large basic units must have special attributes.  However, even a cursory examination suggests that there are many other properties, aside from low elastic moduli, that are also caused by the large size of the component building blocks and that would make such materials distinctive.

\textit{Dissipative matter:}  Because the individual building blocks are large, they each have many low-energy degrees of freedom (\textit{e.g.}, vibrations) that can be easily activated when two particles collide or scrape past one another.  This generates strong dissipation during flow or deformation, either due to frictional interactions or to the inelasticity of collisions.  Kinetic energy, for example due to sound vibrations, is rapidly absorbed into the internal degrees of freedom of the constituents.  Thus granular or colloidal materials are highly damped systems. 

\textit{Disordered matter:}  Because the individual particles are large, it is prohibitively unlikely that any two of them will have precisely the same number of atoms in the same arrangement.  Thus there is disorder in the size and shape of the constituents.  (One counter-example is matter made from viruses, where the DNA regulates the structure precisely.)  Even this small amount of disorder can have dramatic effects.  For example, in a granular material consisting of hard grains, forces will predominantly propagate through the material in paths where slightly greater overlap between particles is present.  Particles that have asperities change the form of the interaction between particles and dominate frictional contacts.  But this is just the beginning.  Because the energy to overcome a barrier between two nearby configurations is small, it is easy for even an initially ordered pack to lose all remnants of long-range order.  Thus much of the physics of dealing with these systems must confront issues of glassiness: slow and non-exponential relaxations, rugged energy landscapes, heterogeneous dynamics (especially near the glass transition), and extra modes of excitation not present in crystalline matter.  

\textit{Non-linear matter:}  Because the component particles are large, it is easy to drive a soft material out of its linear-response regime.  Thus at low forces, it is possible to deform a soft material and even to change its nature from that of a rigid solid into a rearranging fluid.  Descriptions of the excitations within a rigid material must often confront anharmonic effects especially when the system is on the verge of a rearrangement event where non-linear behavior can be dominant.  

\begin{figure}
\includegraphics[width=0.65\columnwidth]{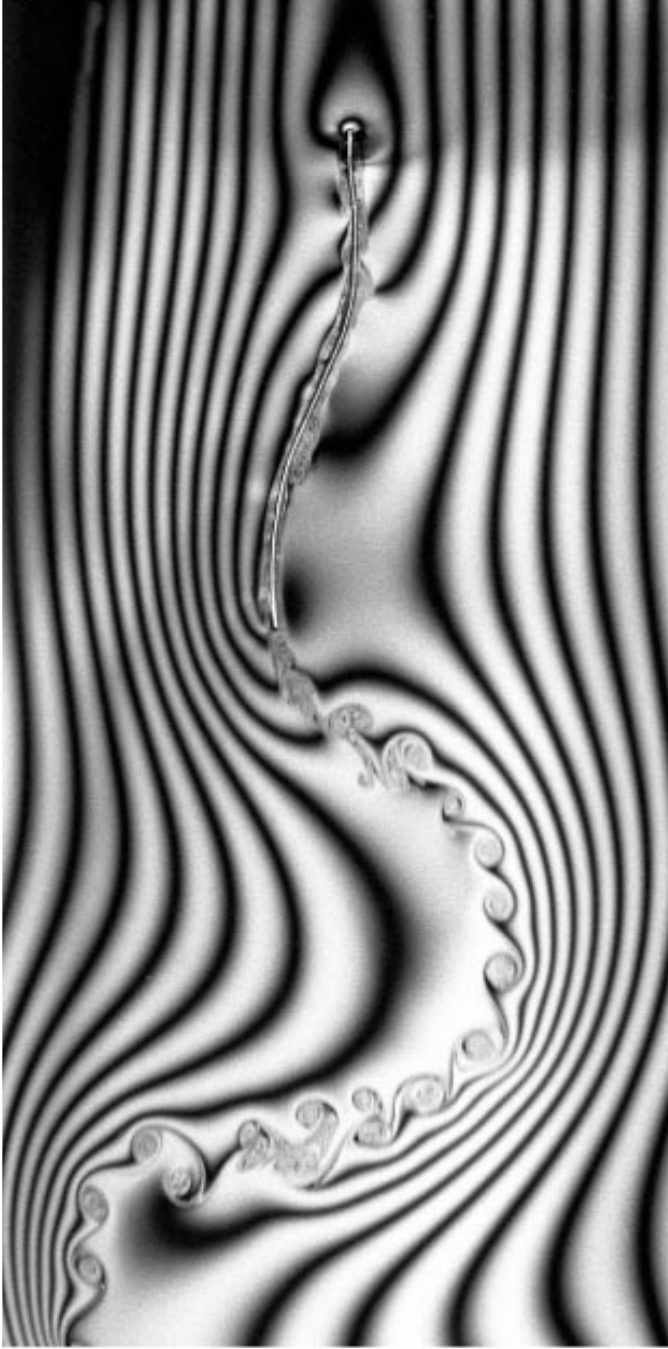}
\caption[]{\label{fig: Figure_SoapflagZhang}
The oscillatory flapping state of a filament in a flowing two-dimensional soap film.  Flow structures, modulated by the moving filament, are advected downstream.  The flow is visualized using interference imaging.  From \cite{Zhang2000}.}
\end{figure}

\textit{Far-from-equilibrium matter:}  For the reasons mentioned above, soft matter is often far from equilibrium.  One class of far-from-equilibrium behavior occurs when continual energy input pushes the system into a dynamic regime where new structures form; relaxation is insufficiently rapid to allow the unstressed state to be an appropriate starting point for describing the characteristic structure and dynamics of the material.  A classic example of such a situation is the turbulent flow of a liquid.  When the boundary of a fluid is flexible, startlingly striking images emerge that illuminate the underlying dynamics.  Figure ~\ref{fig: Figure_SoapflagZhang} \cite{Zhang2000} shows the interaction of a thin filament with a two-dimensional flow.  Pattern formation is often another consequence of energy input overtaking the ability of the medium to relax.

Another class of far-from-equilibrium behavior occurs when thermal energy is insufficient for exploring the relevant phase space.  The glass transition is perhaps the prototypical example of how a system develops such behavior \cite{Ediger1996,Debenedetti2001,Berthier2011}.  In a supercooled liquid, relaxation times become exceeding long as the temperature is lowered.  When these relaxation times exceed the time that any experiment can practically explore, the system falls out of equilibrium and becomes a glass.  While the glass transition occurs in nearly all materials, it plays a special role in polymer science where entanglements can help to suppress the ability to crystallize and freeze the system into far-from-equilibrium configurations.  Many attempts have been made to describe glassy and other far-from-equilibrium materials as having a fictive temperature.  However, it is not at all clear that a single temperature can adequately describe all aspects of the dynamics of a glassy material.

\textit{Thermal and entropic matter:}  Because the energy necessary to deform a soft material is small, room-temperature thermal excitations are often sufficient to rearrange the system.  Indeed, for soft-matter systems, the appropriate energy scale is often given not in units of $eV$, but in units of $k_BT$.  Just by waiting at room temperature under an applied stress, a soft-matter system creeps and deforms. 

For many of the same reasons that temperature plays a significant role, entropic considerations have an increased importance in determining the structure of soft matter.  In many cases, ordered phases of matter are created due to the role of entropy.  Liquid crystals, for example, order into novel phases precisely because creating order in one degree of freedom (such as molecular orientation) can allow many more configurations in others (such as translations).  Understanding the variety of order created out of entropic interactions has given rise to an entire industry.

\textit{Slow matter:}  The diffusion constant for a particle is inversely proportional to its linear size so that the time to diffuse its own length, $L$, increases as $L^3$.  Therefore large particles, such as colloids, diffuse much more slowly than ones of atomic-scale.  Large particles also accelerate more slowly than less massive ones. These are just two reasons why, for example, colloids and granular materials have such long relaxation times.  Even if a material could reach equilibrium, it would take much longer to do so if the comprising particles are large.  

\begin{figure}
\includegraphics[width=0.95\columnwidth]{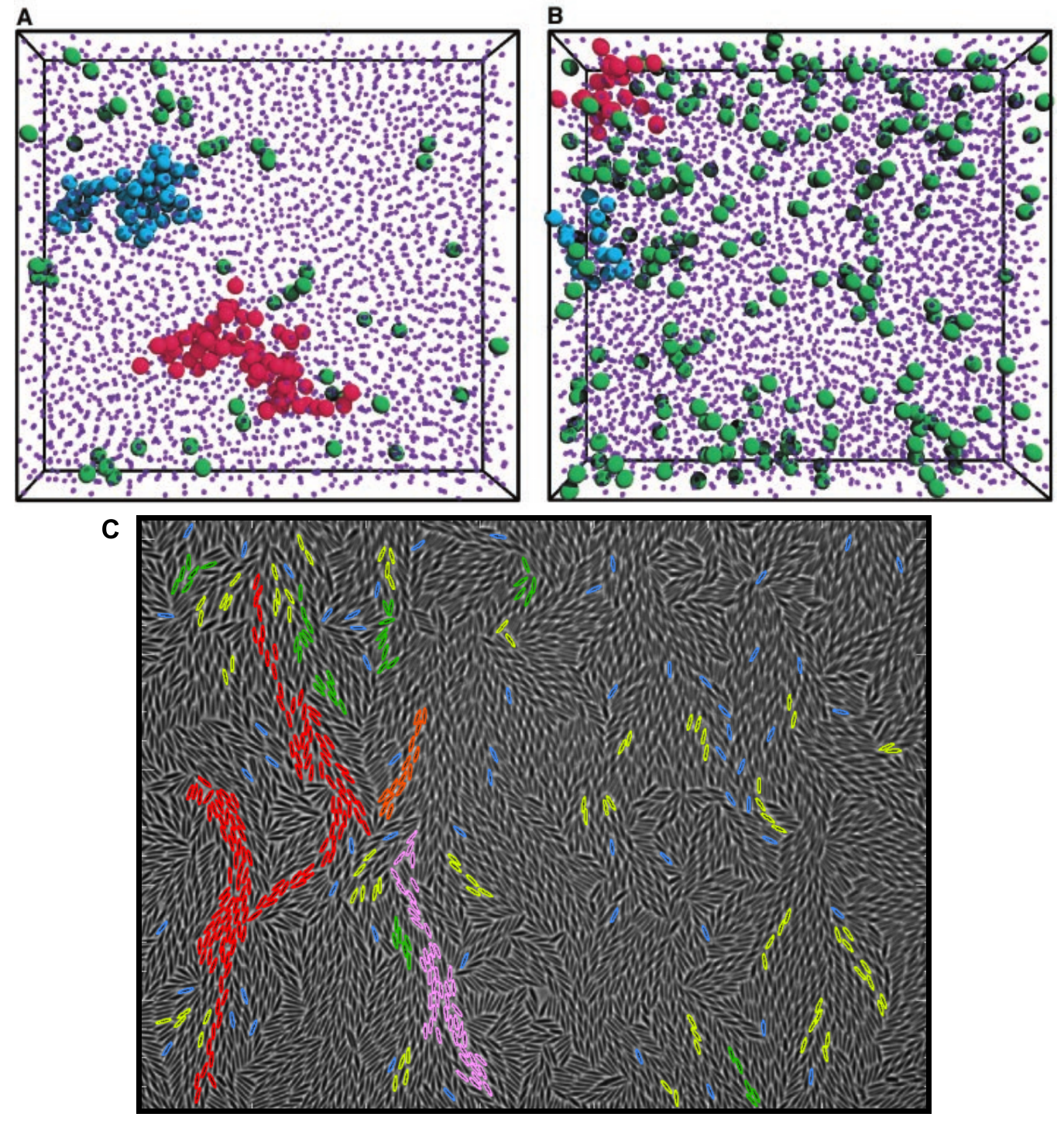}
\caption[]{\label{fig: Figure Glassycolloids}
Fastest particles in (A) a supercooled and (B) a glassy  colloid.  In (A) red (lower central) and blue (upper left) clusters contain $69$ and $50$ particles respectively.  In (B) the largest cluster (red at the upper left corner of the box) contains $21$ particles.  In both (A) and (B) the particles that are not moving as rapidly are shown with smaller size to aid visualization.  From \cite{Weeks2000}.  (C) Motion in a glass composed of ellipsoidal particles.  Both rotational and translational motion can be determined.  Colored (shaded) particles in the image shows the spatial distributions of the fastest $8$\% of translational motions of the ellipsoids.  From \cite{Zheng2011}.}
\end{figure}

\textit{Observable matter:}  One advantage of being slow is that soft matter allows observation of processes that would otherwise be too rapid to track in more conventional situations.  Moreover, large size allows direct visualization of behavior that otherwise could not  be perceived with clarity.  For example, the large individual particles and the slow dynamics of colloidal systems has made them ideal for observing slow relaxation in glass-forming fluids as shown in Fig.~\ref{fig: Figure Glassycolloids} \cite{Weeks2000, Zheng2011} and the vibrational excitations in disordered solids \cite{Chen2010}.  On a different front, because the phenomena appearing in soft-matter systems is often ubiquitous, counter-intuitive and easily observable, these systems are ideal for demonstrating scientific enquiry to a general public and for engaging younger students just starting to think about a career in research.

\textit{Gravity-affected matter:}  Some, but not all, soft-matter systems are naturally far-from-equilibrium because they are affected by Earth's gravity, $g$.  A container filled with sand grains of size $L \approx 1mm$ and density $\rho \approx 1 g /cm^{3}$ is one example where the room-temperature thermal energy, $k_BT$, is approximately $12$ orders of magnitude smaller than $\rho gL^4$, the energy of an elementary rearrangement caused by raising one particle over its neighbor.  As with other far-from-equilibrium phenomena, the question arises: if temperature is no longer sufficient to allow equilibration, how does one describe the state of a sandpile?  What is the ensemble over which one should average in order to obtain a meaningful description? 

\textit{Patterned matter:}  Because soft matter is often far from equilibrium, it is prone to form a wide array of distinct patterns.  As mentioned above, turbulence occurs as a low-viscosity liquid is forced to flow at high speeds.  As the energy cascades from the large length scales where it is injected into the system down to a smaller (Kolmogorov) scale where it is dissipated, new structures are created, such as eddies and entangled vortices, that are not observed in the motionless ground state.  

\begin{figure}
\includegraphics[width=0.95\columnwidth]{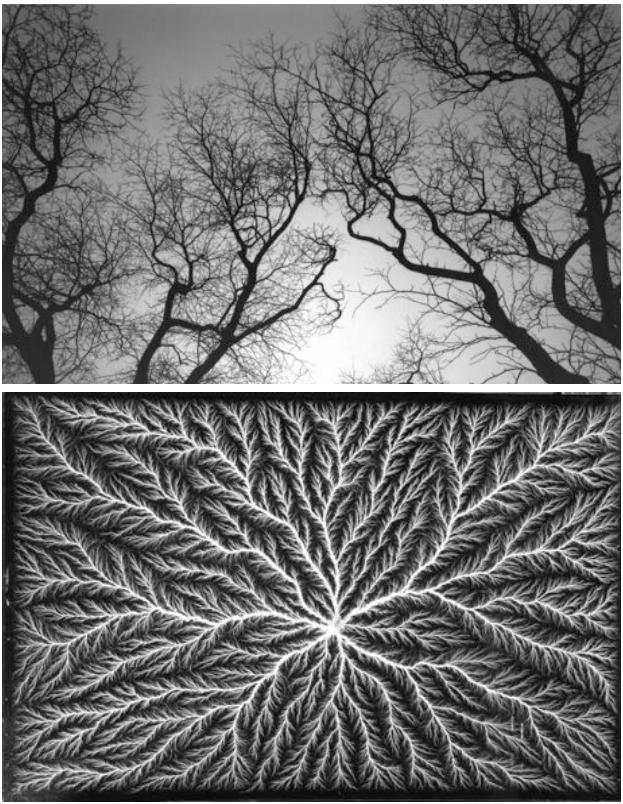}
\caption[]{\label{fig: Dilationsymmetry}
Branching patterns seen in the limbs of trees on top and in dielectric breakdown of a Lichtenberg figure below.}
\end{figure}

Other patterns occur between the constituent phases in a material and display intricate and varied forms in both space and time.  Classic instabilities, such as the viscous-fingering instability \cite{Saffman1958} or the Belousov-Zhabotinsky reaction \cite{Belousov1959, Zhabotinsky1964}, produce mesmerizing and detailed features that progressively unfold in time.  Another common pattern based on dilation symmetry for the penetration of space is shown in Fig.~\ref{fig: Dilationsymmetry}.  Nature appears to be showing off the tricks that it can do! 

\begin{figure}
\includegraphics[width=0.95\columnwidth]{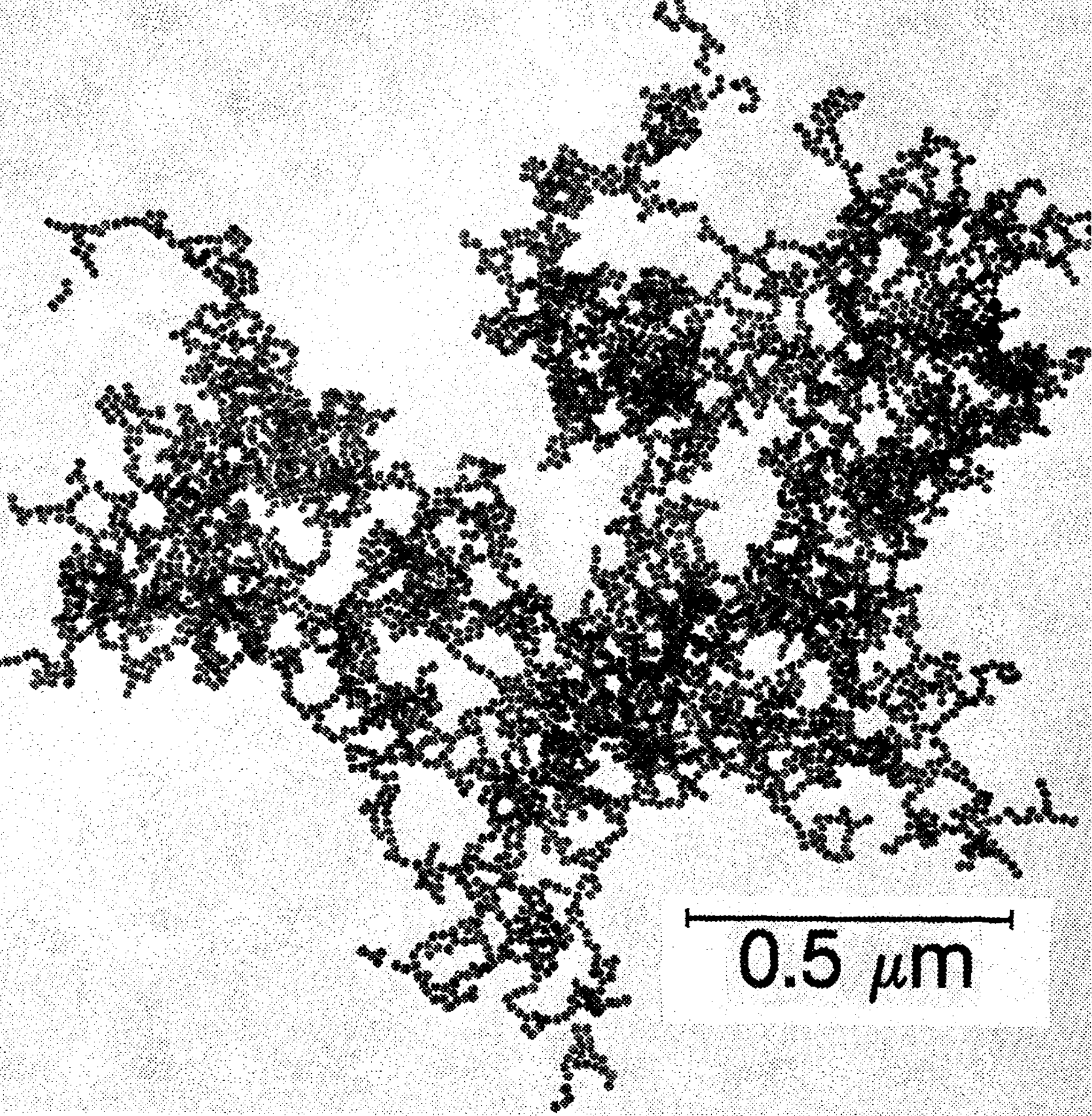}
\caption[]{\label{fig: DLA}
TEM image of gold colloid aggregate showing DLA structure.  From \cite{Weitz1984}.}
\end{figure}

\textit{Non-local matter:}  When patterns form, they cannot always be described by local rules.  Thus, for example, a branch in a diffusion-limited aggregation (DLA) pattern as shown in Fig.~\ref{fig: DLA} can be shielded from further growth by other parts of the interface that are far away when measured along the interface itself  \cite{Witten1981};  when those segments circle back and produce an overhang they effectively prevent other particles from approaching and attaching themselves to the initial protected branch.

\textit{Interfacial elastic matter:}  In hard-matter solids, internal and external interfaces are exceedingly important for controlling the electronic properties of devices.  Interfaces, both internal and external, also play an important role in soft-matter.  Because soft matter is deformable, elastic deformation at a surface is transmitted efficiently into the bulk.  Flow in microfluidic channels and in Hele-Shaw cells are affected by surface topography \cite{ben1985}.  Elastic response and elementary excitations are governed to a large extent by pressure variations at the surface of a jammed solid \cite{Silbert2009}.  Surface tension in a dense suspension governs many aspects of rheology \cite{Brown2012}.  Surfactants (large molecules with a polar head and a hydrophobic tail) change the surface tension at a fluid surface dramatically.  They can produce surface flows and create a myriad array of structures.  Fracture is the creation of an internal surface in a material under stress.  It too is transformed in soft materials into a novel form of material failure.  An effect commonly used in liquid crystals is to use a surface to anchor the orientation of the anisotropic molecules.  This provides one means to control the defect structures in the order parameter.

\textit{Memory-retaining matter:}  Matter that is malleable can be altered by repeated applied stresses.  Because the material would typically be disordered by such treatment, this manipulation can lead to different internal structures that are a signature of how the material has been treated.  That is, the material can retain a \textit{memory} of how it was formed or previously manipulated.  The manner in which such memories can be encoded, read out and erased can vary from system to system.  Sometimes the memory effects can be very subtle.   

\begin{figure}
\includegraphics[width=0.95\columnwidth]{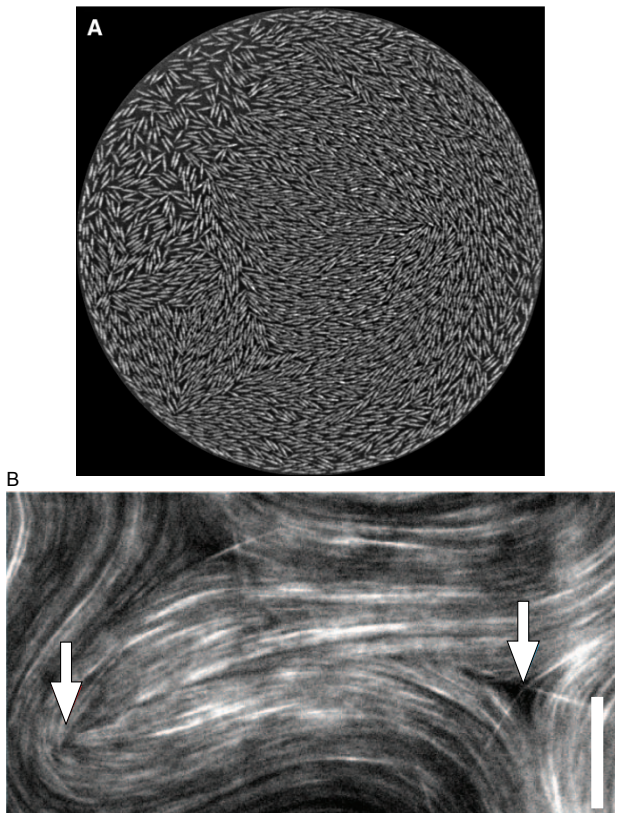}
\caption[]{\label{fig: Figure ActiveLC}
A) Snapshot of nematic order assumed by rods vibrated perpendicular to the plane of the image showing giant number fluctuations.  From \cite{Narayan2007}.  B)  Active liquid crystals exhibiting buckling, folding and internal fracture.  The two arrows indicate the positions of two topological charges associated with the disclination defects in the order parameter.  The scale bar is $20 \mu m$. From \cite{Sanchez2012}.  }
\end{figure}

\textit{Active matter:}  Because the individual ingredients of a soft-matter material are large, it is comparatively easy to have these large entities inject energy at a local scale.  This can be done, for example, by creating different surface activity on different parts of a particle.  When exposed to an external optical beam or when placed in contact with a surrounding fluid, the different ends of the particle can undergo different interactions that will propel them through the solvent in which they are immersed.  When there are a large number of such interactive swimmers, they can produce dramatic swarms of activity and structure \cite{Marchetti2013}.   Two examples of active systems are illustrated in Fig.~\ref{fig: Figure ActiveLC} \cite{Sanchez2012,Narayan2007}.       

\vspace{5mm} 
This list shows an impressive number of ways in which matter created from large component particles is distinct from matter composed of simpler constituents at an atomic scale.  \textit{Softness} is but one of many properties distinguishing these materials from their ``hard-matter'' counterparts.  Indeed, softness is perhaps one of the least interesting attributes listed above.

Each of the characteristics mentioned can allow special functionalities thus creating technological opportunities.  The industry spawned by the new forms of liquid-crystalline order is perhaps the most obvious and well-known example of the immense industrial impact of soft matter.  Likewise, dissipation has been a bane to the efficient processing of materials in industries that manipulate particulate matter.  The cosmetic, paint and food industries rely on the rheology of structured fluids.  Complex flows in polymers lead to dramatic sensitivity of material structure and properties to processing conditions.

The different attributes afforded by soft matter raise important problems in basic science that need to be addressed from multiple perspectives.  Some issues cut across all areas of science; soft matter allows us to address common problems from a fresh perspective.  For example, matter that is far-from-equilibrium is confronted in biology, chemistry, geophysics, astrophysics and nuclear physics.  Even a small increment in our understanding could have enormous ramifications.  Similarly, issues dealing with disorder appear broadly in many scientific endeavors.  Even if not specifically addressing disorder, wherever rough energy landscapes are invoked (\textit{e.g.}, high-energy physics and string theory, biological cell differentiation) much is to be learned by fully understanding the energy landscape underlying glasses and the glass transition.  One can make similar claims about the deep issues confronted in many of the other headings on that list.  In some cases, these materials have provided a novel experimental window for studying common phenomena that could not be easily undertaken in atom-scale systems.  

\subsection{\label{sec:level2}Soft-matter systems and the disciplines that study them}

In addition to the types of behavior displayed, a list of the different classes of systems encompassed by the label ``soft matter'' is also large.  Above, we mentioned a few sample systems: liquid crystals, colloids, foams, granular matter, suspensions and polymers. There are other classes of material that share a strong intellectual base with soft matter.  High on this list would be simple liquids.  Liquids are disordered atomically; they are dissipative (viscosity is important); they are non-linear and can easily form observable patterns; they are soft (but for a reason different from that for soft matter based on large constituent particles) with a vanishing static shear modulus; their interfaces are easily deformed and determine much of a fluid's behavior; they can be pushed far from equilibrium either by constant energy injection, as in turbulent flow, or by super-cooling into a glass phase where thermal energy is insufficient to allow adequate exploration of phase space.  Thus, one home for fluid dynamics is definitely in the soft-matter community.  

Another example that shares common features with soft matter is found in living systems.  Biological matter is soft.  It is often active with energy input on a microscopic level.  Much of biological activity, though purposive, is also seemingly disordered and issues such as crowding and jamming arise.  Living biological matter is the ultimate far-from-equilibrium system.  Soft matter, with its interest in issues tied up with disorder, patterns, structure formation, non-linear and far-from-equilibrium behavior, is a close cousin to biological matter and is arguably the closest intellectual discipline to biology that exists in the physical sciences. 

An emerging activity with a home in the soft-matter community deals with meta-materials.  The aim is to organize structures at a macroscopic level in order to create unique material properties and functionality.  Included in this endeavor are the intellectually rich fields of origami and kirigami based on the folding and tearing of two-dimensional sheets (see \textit{e.g.}, \cite{Chen2016}).  It also includes other approaches to incorporate long-range interactions between local entities or to create global response.  Recent work has seen the incorporation of topological constraints to control excitations. 

Soft-matter research is highly multi-disciplinary.  Because of the enormous range of both materials and attributes, it is not surprising that investigators come from many disciplines including physics, chemistry, materials science, biology, geophysics, applied mathematics in addition to chemical, biological, mechanical, and civil engineering departments.  The range of interests has many overlapping threads.  It is essential to the health of the field to recognize that they all coexist under one roof. 

Soft-matter science provides an excellent tool for outreach and is accessible to a general audience because the phenomena that are investigated are so pertinent to daily life: patterns, food, cosmetics, plastics, modern display technology.  The subject matter adapts itself readily to simple demonstrations, and as such can be exhibited in museums and public lectures.  In addition, the research often creates many stunning and memorable images whose import is easily grasped.

\vspace{5mm} 
There is no ideal way to organize a report on experimental soft matter.  While different \textit{materials} are studied, many \textit{issues} transcend specific materials.  These give intellectual coherence to the field.  The following sections will give an overview of some promising areas of research.  Some topics, such as biologically inspired matter, are not given a section of their own but are referred to throughout the text.  The following sections summarize the discussions that occurred at a workshop held in January 30-31, 2016 devoted to surveying current challenges facing soft-matter experiment.  The participants and discussion leaders are listed in the Appendix. 

\section{\label{sec:level1}Colloids\protect\\}

A colloidal suspension is a mixture of microscopic insoluble particles, between approximately $1 nm$ and $1 \mu m$ in diameter, that are dispersed in another medium such as a solvent fluid.  Perhaps the most commonly encountered example of a colloid is milk.  In a colloidal suspension the particles are too small to settle appreciably under the influence of gravity but are strongly affected by thermal energies.  In other non-colloidal suspensions, the particles are larger; they can sink and are less prone to thermal rearrangements.

Because they do not settle over time, colloidal suspensions can mimic the properties of atomic-scale fluids but at a much larger size and much slower rates as shown, for example, in Fig.~\ref{fig: Figure Glassycolloids}. They thus allow the observation of organizational principles that govern the many-body interactions inherent in strongly interacting matter.  Much previous research has been devoted to understanding simple colloids in equilibrium and at the glass transition \cite{Witten2004,Cates2000,Kleman2003,Hunter2012}.  Colloids have become such a good test system in part due to a variety of tools that have been invented to manipulate and measure their properties with unprecedented precision.  As described in the section on Instrumentation below, these include optical tweezers, diffusion wave spectroscopy (DWS) and confocal microscopy.

New physics can be revealed by adding complexity to the colloids and to their inter-particle interactions and by pushing the systems to be far-from-equilibrium.  The goal is to understand the really complex dynamic interactions of real systems.  These include global collective and emergent phenomena.  Such behavior results from colloidal-scale interactions and can be sensitive to small changes in these interactions.  For example, the field is poised to address the spontaneous assembly (or disassembly) of structures, the emergence (or averaging out) of large-scale fluctuations.  Understanding some of these issues would be important for material processing. 

One forefront issue is to use colloidal systems to study the onset of non-equilibrium behavior.  While, the small size of colloids often allows them to reach equilibrium, the relaxation times are sufficiently long that the system cannot always keep up with the forcing.  For example, slow compression can push a colloidal system into a glassy or jammed phase where the system has fallen out of equilibrium with the thermal bath.  These glassy systems, with radial inter-particle interactions, provide a minimal model for systems like metallic glasses and are an excellent model system with which to study the transition from a fluid to a rigid solid and the transition from a thermal to an athermal system, topics discussed elsewhere in this report.  A fascinating aspect would be to use colloids as a tool to gain insight into non-equilibrium behavior.  Can these systems reveal dynamics in unexpected experimental phenomena?  

Colloidal aggregates are often disordered.  While sometimes a bane, disorder can also provide advantages: it can make new material properties accessible that would otherwise be difficult to obtain in ordered matter.  For example, disorder can provide fault-tolerant tools.  Also, because of disorder, a given bond can have widely varying contributions to different global properties \cite{Goodrich2015}.  Can novel materials be made from disordered colloidal aggregates that use this property?    

Another avenue to be pursued is to push the study of colloidal materials into new and extreme regimes.  For example, these systems could be pushed to higher driving rates and into new physical environments.  Their dynamics would need to be studied at much faster time scales.  Colloidal suspensions can also be driven by external forcing or can be made ``active'' so that the individual particles have their own energy source.  This suggests qualitatively new capabilities for how materials self organize.  For example, it can provide an experimental platform to study flocking or swarming.  Likewise nanoparticles can display directional motion in response to external fields.  

Biological matter has many mechanisms that exquisitely accomplish and control specific tasks.  For example, a protein can interact with another molecule at one point on its surface which then allows the binding of a second molecule at a distant location.  Such ``allosteric'' interactions are an important way in which protein activity is regulated \cite{Ribeiro2016}.  Colloidal systems may present an opportunity to study allostery in a physical system and examine what aspects of structure allow specific and localized long-range interactions.

Another bio-inspired example is to use \textit{E. coli} as an inspiration for designing colloids (which, in analogy, we call ``e-colloids'') that would report on chemical interactions within a material and react to them in a specified way.  A major challenge is to design a more general chemistry that could engineer colloids with selected surface and bulk properties in order to react to, and report on, a variety of interactions.  For example, a tissue spanning two distinct (immiscible) fluid phases, taking information from one phase to cause a reaction in the other, could form a natural platform for decision-making structures.  This also offers the possibility of remote control or robotic applications.  The colloids can act as sensors and amplify the signal at larger scales.  Another goal would be to design colloids that respond only when the collective arrangement of the particles is in a specific structure.  Such reporter e-colloids would generalize one of the most amazing mechanisms in the biological world and would open up new ways of designing smart and programmable materials.  

In closing this section, one should mention the forefront issue of designing materials with specified rheological properties.  For example, it would be useful to have materials with a temperature-independent viscosity or materials that behave in desired ways under extreme shear.  These are examples of an inverse problem; one starts with a desired property and then attempts to figure out how to make that property occur.  This aspect of materials science necessarily requires a strong interaction between theory, computation and experiment.    

\section{\label{sec:level1}Granular Materials, (non-Colloidal) Suspensions, Emulsions and Foams\protect\\}

Granular physics is central to many aspects of soft-matter research.  This subfield concerns materials composed of macroscopic ``grains'', which are usually solid particles but also may be liquid droplets as in emulsions or gas bubbles as in foams \cite{Jaeger1996,Halsey2003,Duran2012}.  By contrast with the colloidal regime, the particles are large enough that thermal motion is usually negligible.  Grain-grain interactions are often dominated by contact forces, but may also be mediated by an interstitial fluid or by electrostatic interactions. There are strong scientific connections with other areas highlighted in this report, including colloids, mechanics, rheology, glasses and jamming, simulations and big data, pattern formation, and packing.  Active matter, human crowds, and traffic also have a granular nature but involve more complex ``grains'' with ``social'' interactions. 

It is worth emphasizing at the outset the enormous range of size and conditions where granular matter appears.  It can be found as sand on beaches, as screes on mountain sides, as icebergs floating off the edge of glaciers, and, at the celestial scale, as planetary rings and asteroids.  It is also found ubiquitously in industrial processes from the pharmaceutical to the civil-construction industries.  Understanding the science of granular matter is important for many societal needs with implications for soil remediation, mining and resource extraction, carbon sequestration, erosion, sedimentation, siltation, and water purification. 

Granular materials exhibit four prototypical properties: they are (i) disordered and heterogeneous, (ii) highly dissipative, (iii) far-from-equilibrium with (iv) strongly nonlinear response to applied forces.  Because there is no understanding of what ensemble can be used to obtain average properties, there is no adequate statistical mechanics of a granular material.  The usual tools and concepts we rely upon for crystalline materials, fluids, and even most other forms of soft matter, simply do not apply.  Friction, important for both statics and dynamics, is poorly understood.  Except in the most idealized situations, it is not understood how friction (and dissipation generally) changes the way a material reacts to external perturbations.  It has thus proven uncommonly difficult to understand and manipulate the behavior of a large collection of seemingly simple objects such as sand grains.  The emergent behavior can be surprising, often with no parallel in other forms of matter as has been emphasized in various reviews. 

In many cases, we encounter granular material in a static, but far-from-equilibrium, configuration.  This is an example of a jammed state of matter.  Because there are external stresses (gravity or applied stress) and because thermal energy is too negligible to create a rearrangement event, the system is doomed to be stuck in an amorphous state that is far from its lowest-energy ground state.  The idea of jamming has been a productive way to think about other far-from-equilibrium systems \cite{Liu2010}.  While much effort has been devoted to studying jamming in the simplest of cases, current research and future opportunities involve focusing on more complicated and realistic systems with more general particle shape and interactions.  Recent attention has focused on understanding shear jamming, the flow of frictional particles under conditions of uniform shear \cite{Bi2011}.  New issues arise when one considers spatial heterogeneities and boundary effects.  The kinetics of jamming, involving the time evolution of a flowing system, is a broad new challenge.

\begin{figure}
\includegraphics[width=0.95\columnwidth]{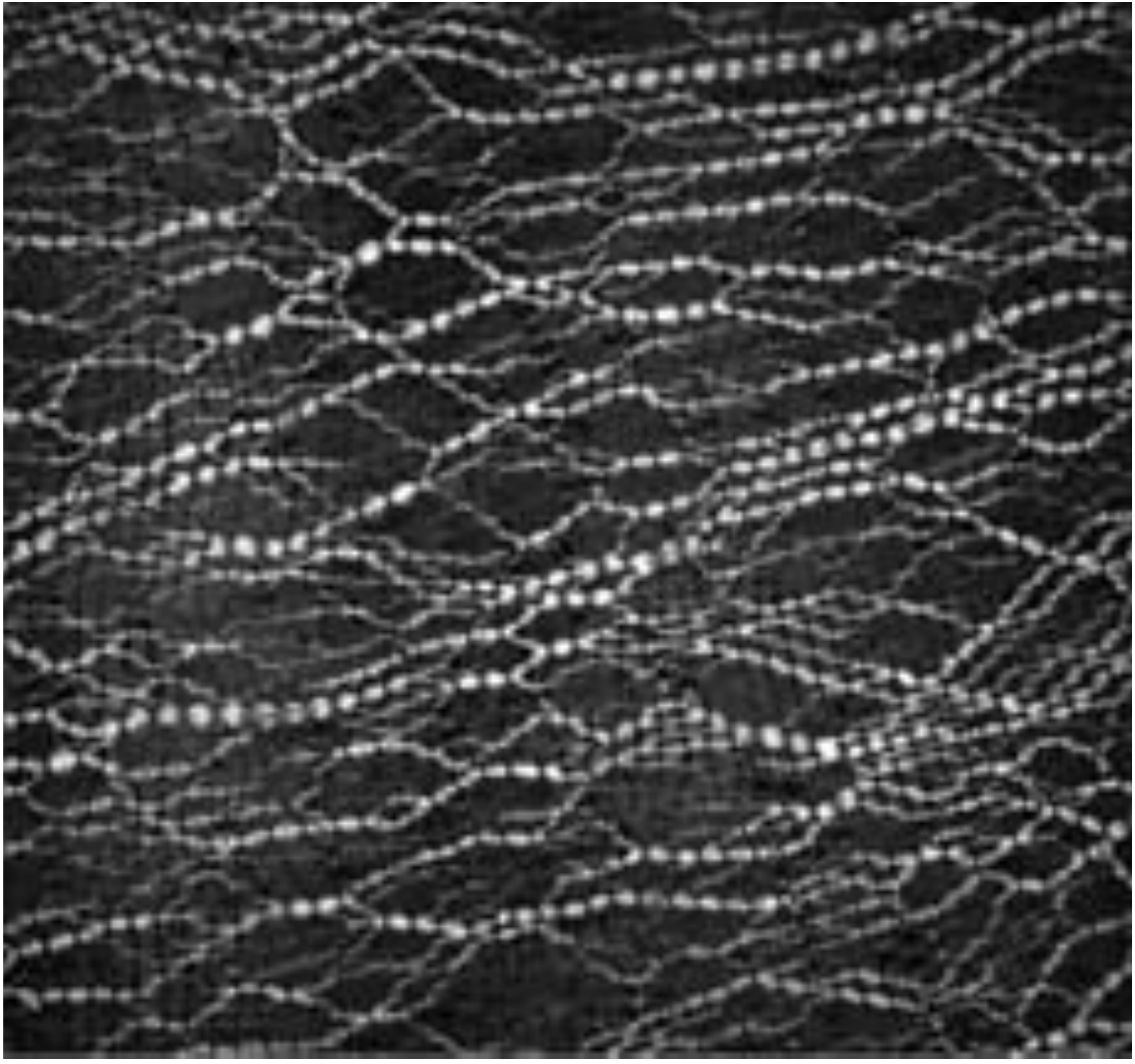}
\caption[]{\label{fig: Figure Forcechains}
Force chains in a two-dimensional granular material that is compressed uniaxially.   The material is placed between two crossed polarizers.  When under stress, the granular particles become birefringent and rotate the polarization of the transmitted light. The particles that are under stress are thus made visible so that the stress paths, from one side of the material to the other, are made apparent.  From \cite{majmudar2005}.}
\end{figure}

In studying the mechanics of a dry (or indeed, even a wet) granular medium, one can ask how microstructure, \textit{e.g.}, particle shape or roughness, affects the dynamics of flow or the force transmission through a static pile.  At a larger scale, a widely discussed, but  unresolved, issue is the role of mesostructure (structure that can be identified at an intermediate scale between the particle size and the system dimension).  For example, Fig.~\ref{fig: Figure Forcechains} shows the existence of force chains in a two-dimensional granular material under stress.  How does such mesostructure affect the material properties and can it be manipulated to control a material's behavior?

Even a seemingly stationary granular material has complicated dynamics if it is under an applied stress.  For example, it can show sub-threshold creep where local particle rearrangements allow small reductions in the stress.  Thus, one needs to understand the nature of the linear regime for the different elastic moduli.  The material can also give way in a large-scale avalanche  \textendash~ a prototypical catastrophic failure mode of a material along a slip plane.  The failure mechanisms depend on the formation history of the pile and how it was packed.  The essence of avalanche behavior can perhaps be captured in the collapse of simple experimental piles of sand or in simplified cellular-automata models.  Clearly, however, richer sets of behavior can be found in real-world situations encountered, for example, in geophysical phenomena (and lab-scale counterparts).  

\begin{figure}
\includegraphics[width=0.95\columnwidth]{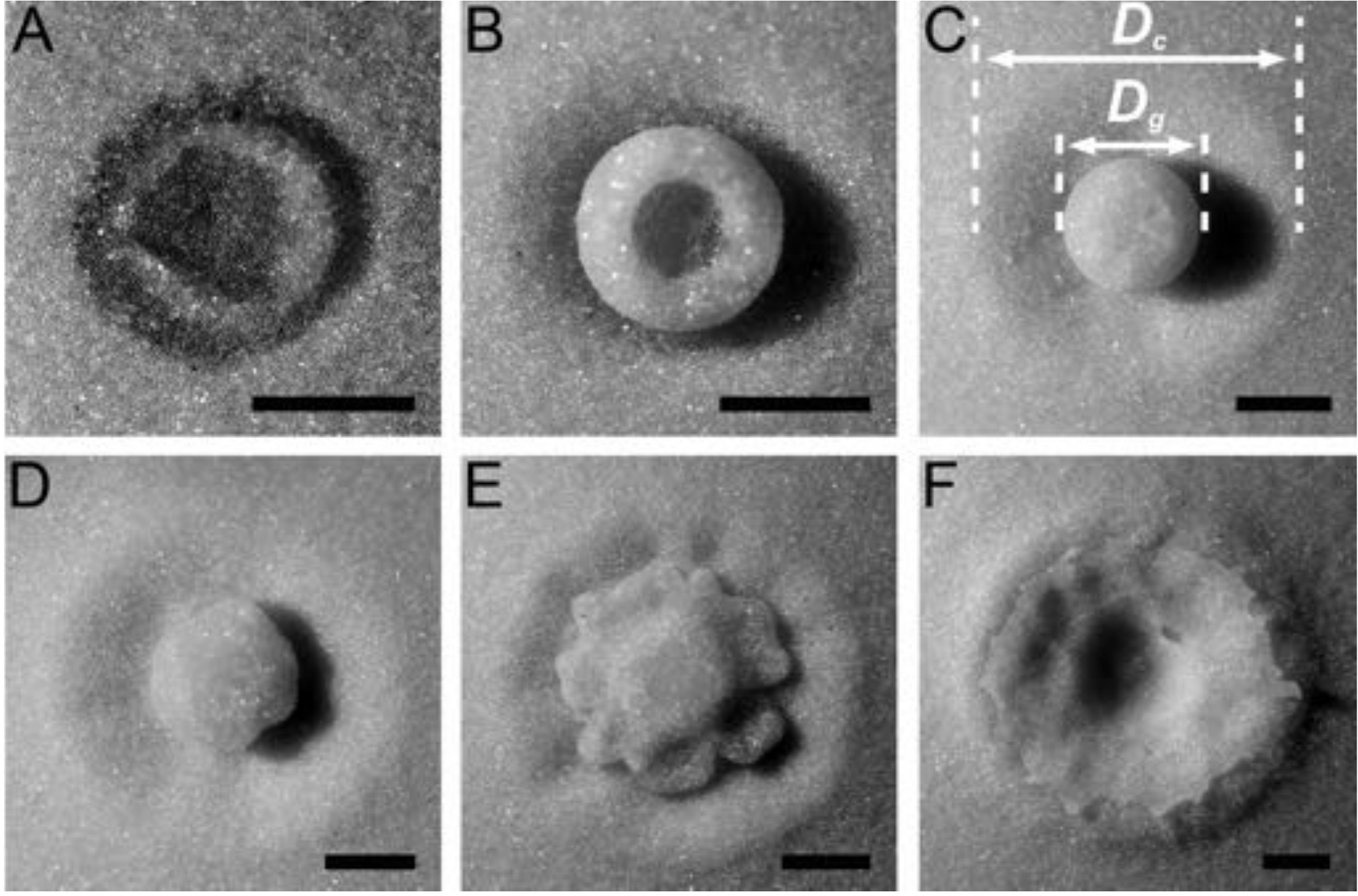}
\caption[]{\label{fig: Figure_GranulardropsCheng}
Liquid drop impact craters from a 3.1 mm water drop. Images show ring-shaped, solid pellet-shaped, asymmetric residues as impact parameters are varied.  From \cite{Zhao2015}.}
\end{figure}

Just as Newtonian fluids routinely display a great range of patterns, their complex cousins \textendash~ granular materials with their non-linear and dissipative interactions \textendash~ likewise demonstrate a wealth of structures (such as oscillons \cite{Umbanhowar1996}, localized excitations in thin granular layers) when subjected to vibration.  
When a granular material flows, its flow profile is highly heterogeneous and the material has a non-local rheology.  Many remarkable phenomena take place such as the segregation of grains according to their physical properties such as size, roughness or density.  When a projectile suddenly impacts a granular bed, as would occur in a meteor hitting the earth, the material gives way and flows in distinctive patterns as shown in Fig.~\ref{fig: Figure_GranulardropsCheng}.    

One phenomenon associated with flowing dry grains is the build-up of static electricity as individual particles become charged and aggregate into larger clumps.  This can occur even if the particles interact with others made of the same material.  Such charging is ubiquitous, important for many industrial applications, puzzling and poorly understood \cite{Pahtz2010}.  In addition, particle shape, heterogeneity, bonding (including van der Waals forces, capillary bridges, or chemical bonds) and many-body elasticity are important for understanding the nature of granular flow.  With all of these interactions, it is a challenge to design the properties at the grain level in order to control a material's behavior.

\begin{figure}
\includegraphics[width=0.95\columnwidth]{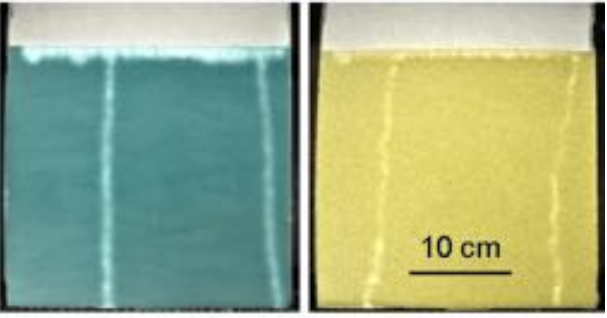}
\caption[]{\label{fig: Figure Waterchannels}
Channeling of water in dry hydrophilic sandy soils at different grain diameters.  In the left panel, the grain diameter is $0.5$ mm; in the right panel it is $1.0$ mm.  From \cite{Wei2014}.}
\end{figure}

A vast and industrially important area has to do with granular suspensions in which the grains are dispersed in a fluid.  Such fluids have a complex rheology with shear-thinning and shear-thickening regimes.  Cornstarch in water (oobleck) is a particularly dramatic example of the latter regime; a person can run across a vat of the material but will sink if she pauses at any point.  The study of this extreme shear thickening has required the development of new experimental tools.  Other aspects of granular suspensions that are currently being investigated have to do with sedimentation, clogging and erosion.  An example in Fig.~\ref{fig: Figure Waterchannels} shows the morphology of water channeling in sandy soils \cite{Wei2014}.

\begin{figure}
\includegraphics[width=0.6\columnwidth]{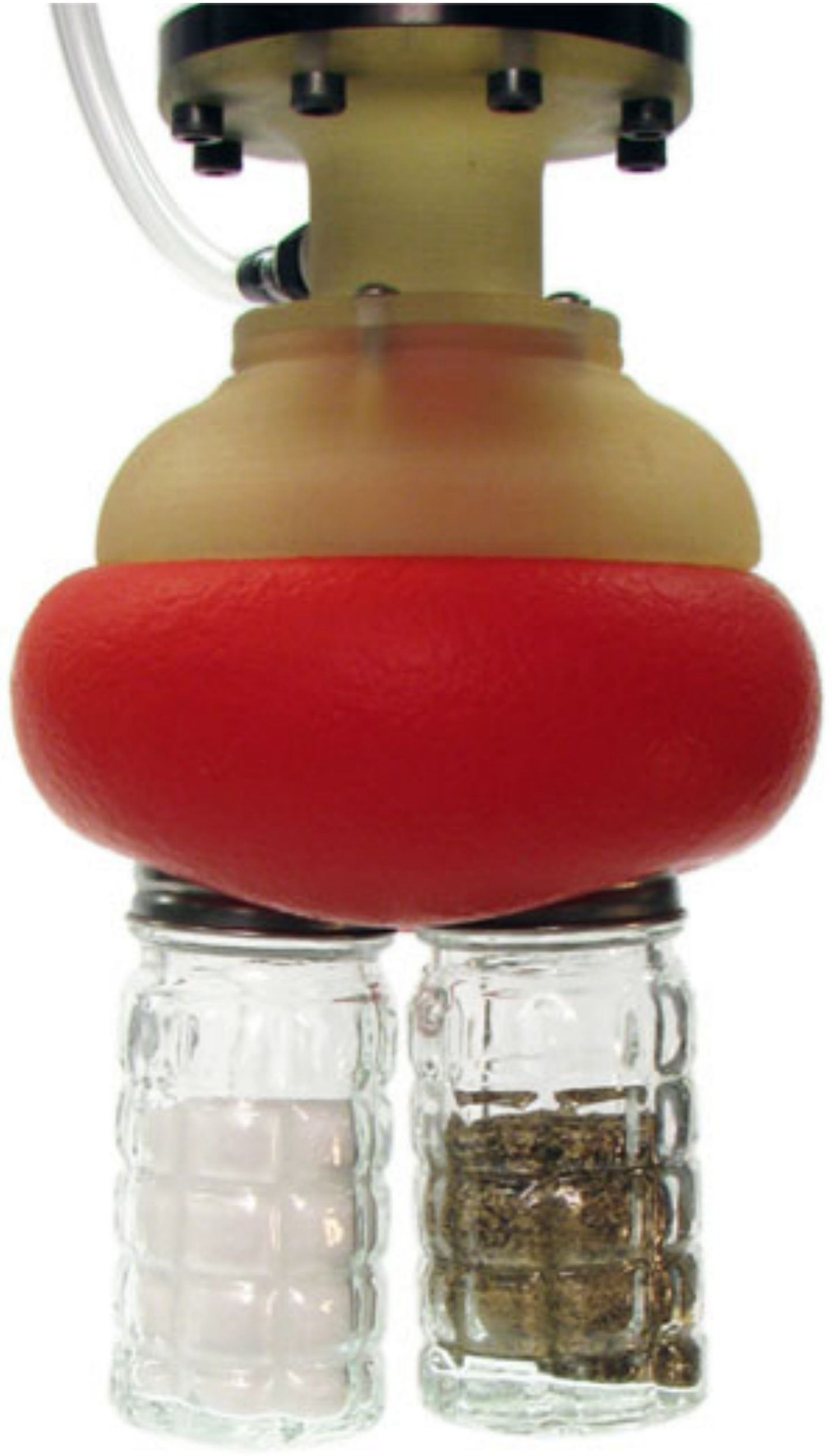}
\caption[]{\label{fig: Figure Granulargripper}
Universal grippers based on jamming in granular material.  A rubber membrane, containing a granular material, can be deformed so that its shape conforms to that of an object to be manipulated.  The membrane can be evacuated so that the granular material becomes rigid due to the external air pressure.  This produces a universal gripper that can lift a large variety of objects with different shapes.  Here the gripper is shown picking up a salt and pepper shaker. From \cite{Amend2012}.
}
\end{figure}

Active materials are emphasized throughout this report.  Granular matter is an arena where the challenge of creating such materials with smart, shape-changing and reporting particles can be found.  Soft robotics based on granular materials has demonstrated exceptional success as shown in Fig.~\ref{fig: Figure Granulargripper}.  It provides a promising new way of using jamming to control motion and rigidity.   

There are an evolving set of tools that are being used to study granular material in all its forms.  These include magnetic resonance imaging, X-ray tomography, high-speed video and ultrasound that reveal structure, confocal microscopy that can measure contact forces, big-data, machine-learning and evolutionary algorithms that can design, analyze and identify novel phenomena in experiments.  Dusty plasmas now offer a promising way to study interactions in granular matter in new ways.

\section{\label{sec:level1}Glasses and Jamming\protect\\}

Glass formation and jamming occurs when a material that can flow loses that ability as either the temperature, pressure, density or some other control parameter is varied \cite{Ediger1996,Debenedetti2001,Berthier2011}.  It differs from crystallization, the more easily understood but not necessarily more commonly found form of rigidity formation, which occurs when a liquid, cooled below its freezing temperature, nucleates crystallites that then grow to encompass the entire system.  The issues associated with rigidity in amorphous solids are so common and relevant that the study of all glass-forming atomic and molecular systems is aptly covered by the umbrella of soft-matter science.  

While window glass is the most common example of a glass, nearly all liquids can be made glassy if only they are cooled rapidly enough.  (Counter-examples are ordinary water and, of course, liquid helium at atmospheric pressure.)  There are metallic, semi-conducting, polymeric and insulating glasses.  On larger scales, glass formation has been studied in colloidal fluids, and jamming of course occurs in suspensions and macroscopic samples such as foams and sandpiles.  It has been much debated how the physics of the glass transition overlaps with the physics of jamming.  Is the glass transition universal (the same for a wide array of materials) or is it system-dependent so that the details of bonding (\textit{i.e.}, covalent versus ionic versus hydrogen bonding) matter?  More provocatively, one can ask whether the glass transition in an associated liquid is the same as the freezing of the dynamics in a colloidal sample or the yielding transition in a granular material.  Such questions were the genesis of the jamming phase diagram \cite{Liu2010} where it was postulated that understanding \textit{e.g.}, glasses would be relevant to other types of amorphous matter.  Dynamic heterogeneities have been studied in many materials as the system dynamics becomes arrested \cite{Berthier2011Book}.  

As an ordinary liquid is supercooled, the dynamics slow down until the time for relaxation becomes much longer than any observer is prepared to wait.  That is, the material falls out of equilibrium when the thermal energy in the sample is insufficient for the system to explore adequqtely the phase space in the accessible experimental time.  Glass formation is the prototypical example of how a system can fall out of equilibrium, not because it is being constantly driven, but because its thermal energy is too small.  Thus when studying a glass, one should be aware that this state was reached, not by remaining in (quasi-) equilibrium while cooling through a transition, but by the system being forced to fall from equilibrium.  In that sense, experimental low-temperature glasses are similar to the jammed state of matter.

Unlike crystals, glasses exist in an incredibly large number of nearly equivalent states in an exceptionally ramified, complex energy landscape.  Dealing with so many relevant energy minima is one of the central problems of statistical physics.  As such, understanding glassy behavior can open a window on many other physical problems, ranging from cosmology and geology to chemistry and materials science, that deal with rough energy landscapes.  It is useful to divide the studies of glasses into two categories (i) those that study the glass transition and (ii) those that study the properties of the glass itself and investigate how it is different from a crystal \cite{Anderson2012}.

One question that has plagued the glass-transition literature for decades is whether there is a true thermodynamic phase transition that separates the liquid and the glassy states as the temperature is lowered.  From experiments alone, it has been impossible to answer this seemingly simple question.  One finds that as the temperature is lowered, relaxation times and viscosity rapidly increase by over $17$ decades in a narrow temperature range \cite{Ediger1996}; but it is impossible to tell if they diverge because no matter how patient the experimenter, her experiment invariably falls out of equilibrium at temperatures that are tens of degrees away from where the divergence would appear to exist.  All one can do is extrapolate into this low-temperature, long-relaxation time region.  Simulations on large particulate systems likewise are cut off by this inability to go to long times (although recent simulations have obtained systems with time scales that are comparable to those in experiment \cite{Berthier2016}).  

When placed under stress, a glass will creep, yield and flow.  It is often attractive to think of these processes in terms of the motion of defects.  While the concept of a defect is clear in a crystalline material, an important current issue is to understand what constitutes a defect in an amorphous solid.  It is far from obvious what those defects might be in a glass, granular pile or other amorphous material.  This has led to a body of research attempting to identify from the local structure the location of soft spots where a particulate material is most likely to undergo a rearrangement \cite{Schoenholz2016}.  Much remains to be done to understand soft-spot dynamics and how such ``defects'' interact with one another.   

Colloidal suspensions can be a way to study the physics of rigidity formation as a liquid state is cooled (or compressed) as would occur when a glass is formed out of a supercooled liquid.  As shown in Fig.~\ref{fig: Figure Glassycolloids}, colloidal systems have allowed the visualization of local rearranging clusters that appear as a fluid loses its ability to flow freely \cite{Weeks2000}.  In recent years, colloidal systems have been the most highly exploited experimental system for testing the ideas that have emerged from computational studies of jamming \cite{Liu2010}.  By studying the position and dynamics of individual particles, it has been possible to observe the non-monotonic correlation function as a function of compression, to measure the spatial properties of localized normal modes, or to observe soft spots, the sites most likely to fail as the material is stressed \cite{Zhang2009,Caswell2013,Chen2011}. 

New challenges await in this area. These have to do with studying how a particle's (non-spherical) shape affects jamming and glass formation and how more complicated inter-particle interactions change the transition and the nature of the dynamics in the rigid phase.  For example, studies of colloids that have long-range or three-body interactions, have non-spherical shapes, or have frictional contacts can produce considerably different physics from the simple frictionless spheres with finite-range repulsive interactions most often studied on computers.  It would be of particular importance to understand the role of a small temperature on the jamming scenario.  While some rearrangements would be thermally allowed, they are far from sufficient to bring the entire system to equilibrium.  It is also important to consider what other axes are relevant to the jamming phase diagram or whether the three canonical ones (temperature, inverse density and shear stress) are sufficient.  

As in so many of the other areas covered in this report, the possibility of creating glasses and jammed matter out of active particles is the start of an entirely new area of research.  Active swimmers have correlated motion that can lead to crowding and slow dynamics.  Are the properties of such systems the same as the conventional amorphous particle counterparts or are there new features that emerge?  Are the flows in such active materials related to the excitations of their static counterparts and can one gain intuition into biological crowding by its analogy with jammed systems \cite{henkes2011}?     

\section{\label{sec:level1}Packing\protect\\}

A material's properties, such as whether it is jammed or able to flow, are obviously dependent on how its constituent particles are assembled.  When objects pack, one would like to be able to predict how they will assemble and what the packing density and overall properties of the material will be.  One property of paramount importance for many applications is the threshold to material failure from different applied stresses.  To make headway, one needs to know what inputs about the particles and about the preparation conditions are necessary to produce accurate predictions.  

The importance of this problem becomes clear when one considers the canonical example of two containers filled with granular matter with different preparation or processing histories.  As originally poured, a packing of spheres will have one density but after just a few taps or shakes (or by more gentle sedimentation) the grains settle to a density that is over $10$\% greater than its original value in the rapidly poured system \cite{Onoda1990}.  Particles with different shapes will have very different densities \cite{Glotzer2007}.  

There are also more subtle effects; packings of spherical grains that were prepared by different tapping protocols to have the same density can be told apart by the way in which they respond to further perturbations \cite{Josserand2000}.  This is a form of material memory that is hidden in the packing structure.  In another example, the forces measured at different points along the bottom surface of a sandpile (due to the weight of the sand above it) varies with preparation conditions \cite{Wittmer1996,Bouchaud2001, Geng2001}.  Can small perturbations (such as from thermal expansion) disrupt the subtle structure of such packings?

It has proven notoriously difficult to characterize experimental packings.  It requires tools that can acquire dynamic images of static and actively driven packings over a large range of scales deep into the bulk.  Magnetic resonance imaging \cite{Nakagawa1993} or X-ray tomography has severe limitations in terms of resolution, speed, and size.  Ultrasonic imaging \cite{Han2016} has issues with contrast but may be useful in some experimental situations. When one goes to nanoscale particles, there are challenges for obtaining real-time imaging.    

\begin{figure}
\includegraphics[width=0.95\columnwidth]{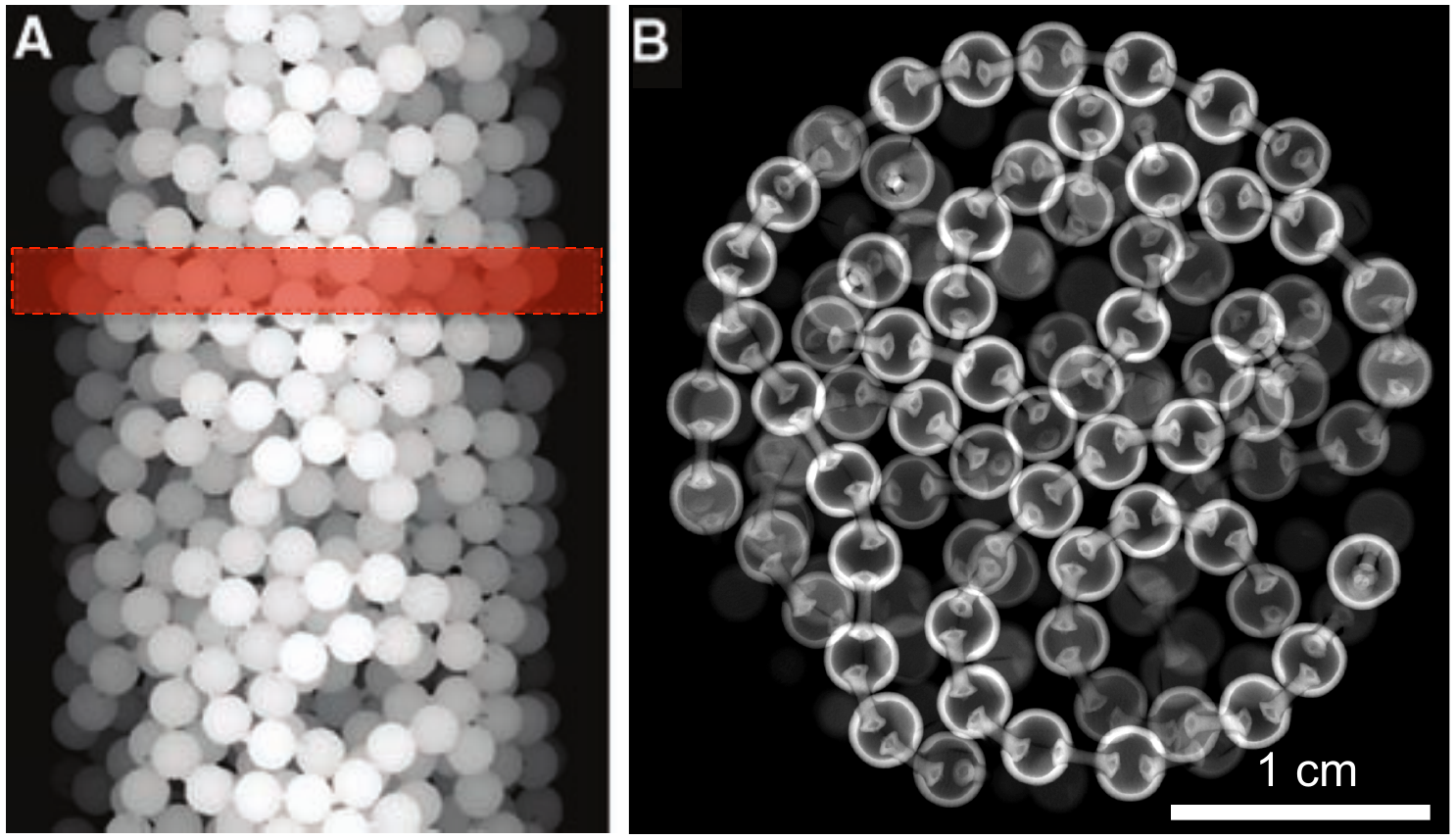}
\caption[]{\label{fig: BallchainZou}
(A)  X-ray tomography image of a ``granular polymer'' consisting of many equal-length segments of a connected chain of spheres.  (B) Close-up image of red (shaded) region in (A).  The connecting links are visible from which it is possible to determine which spheres are attached together in each chain.   From \cite{Zou2009}.
}
\end{figure}

To create reproducible packings, one must control the constituent particles' shape and size.  Moreover, friction certainly plays a central role; frictionless particles simulated on a computer do not display nearly the range of density variation that is found in laboratory packings.  It is essential that all the interactions between objects (\textit{e.g.}, friction, cohesion, charging) be understood.  Monodisperse spheres have been fairly well studied.  However, there is great latitude to create more exotic ingredients.  It is now possible to create designer particles of different shapes by using various chemical syntheses or, at a macroscopic scales, by using 3D printing.  One can have particles that are polydisperse, non-spherical, polymeric or with concave surfaces.  Likewise, boundary conditions, finite-size effects, external fields (\textit{e.g.}, gravity) are important.  An example is the polymer-chain packing visualized with X-ray tomography shown in Fig.~\ref{fig: BallchainZou}.  One goal is to be able to predict reliably the structure, density and failure properties of such a packing.

For many purposes, one wants packings with tailored properties such as extreme loose or close packing.  Thus one wants to enhance or frustrate high packing densities.  The role of packing becomes even more important when one uses shapes that are more complicated than simple spheres.  One goal has been to find the particle shape that gives the highest-density packing \cite{Donev2004,Torquato2010,Roth2016}.  Another goal is to determine the limit of material parameters that can be obtained from different protocols.  In the inverse problem, one wants to predict what shape particles should have to create a packing with a particular property.  One successful example used an evolutionary algorithm to predict a particle shape that matched a particular desired rheological feature \cite{Miskin2013}.   

\section{\label{sec:level1}Liquid Crystals\protect\\}

Liquid-crystals exhibit some properties that resemble those of a liquid while others that resemble those of a solid.  This behavior stems from the fact that the large constituent molecules in a liquid crystal are anisotropic and are arranged with an intermediate degree of order in some directions while they appear amorphous in others.  Thus liquid crystals may flow while still retaining some molecular order.  There are many different types of liquid crystals and, because the order in these materials is highly anisotropic, many liquid crystals have unusual and highly useful optical properties that can be manipulated to great effect in display technology.  This has led to a $> \$100$ billion/year industry \cite{Kent2012} and many new phases of liquid crystals are discovered each year. 
With this abundance of opportunity and new material, it is not surprising that many fundamental issues need to be addressed.  

The symmetries and director order parameters that describe liquid crystals lead to a variety of topological defects that mediate the material's structure.  For example, when the director is pinned at two points with incompatible alignments, a topological defect may be required to mediate the transition between the two regions of space.  Defect lines can even be tied into knots \cite{Tkalec2011}.  Experiment is essential for uncovering many of these subtle and intriguing structures and one can visualize such topologically complex fields by making use of the optical birefringence of these anisotropic materials. 

\begin{figure}
\includegraphics[width=0.95\columnwidth]{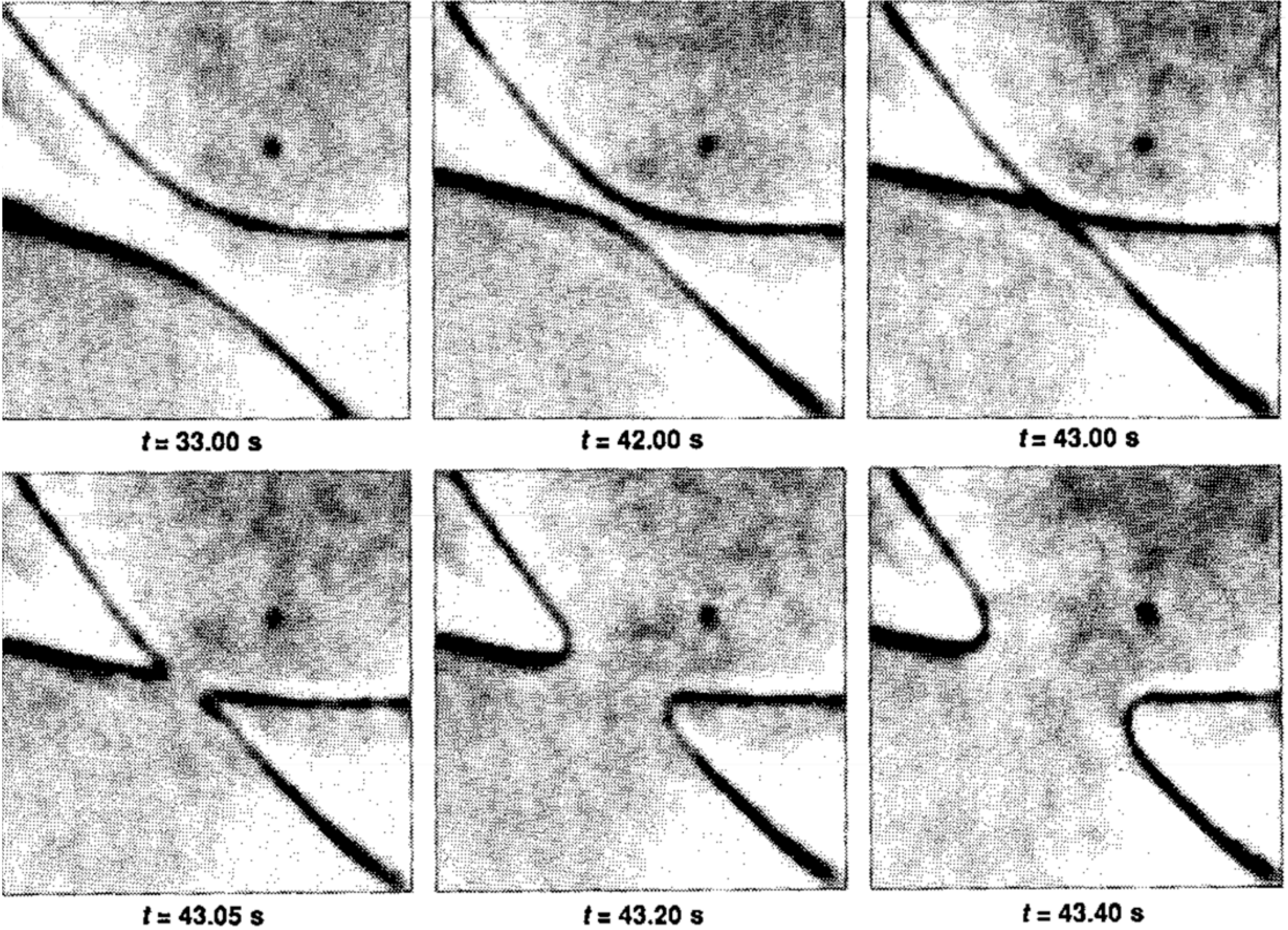}
\caption[]{\label{fig: Figure Cosmicstringdefects}
Two type-${\frac 1 2}$ strings in a liquid crystal crossing each other and reconnecting.  These defects have been taken as analogs of cosmic strings in the early universe.  From \cite{Chuang1991}.}
\end{figure}

Liquid crystals also provide a unique platform for answering fundamental questions about how nature can create novel forms of order, not only in condensed matter but also in more general contexts.  Indeed, some of the defects that have been discovered in liquid crystals, such as those in Fig.~\ref{fig: Figure Cosmicstringdefects}, have been cited as similar to those that appear in the fabric of the early universe \cite{Chuang1991}.  Thus liquid crystals not only have had important technological applications but also have provided an exceptional resource for posing and investigating fundamental questions about collective behavior.

A recurrent theme throughout this report, and of particular importance here, is the structure and dynamics when a material is far from equilibrium.  One can investigate the classification of non-equilibrium states and study the kinetics of defect motion.  There are orders of magnitude in the amplification of response imparted by the long-range liquid-crystal organization.  Moreover, the response can often be rapid.  Although this amplification is at the core of many applications, its origin is not well understood.  

Another aspect occurs if one combines the polarity and flow of liquid crystals with activity, so that individual particles have their own energy source.  In such cases, new regimes of behavior and organization can be expected as seen in Fig.~\ref{fig: Figure ActiveLC}.  The organization of such active liquid crystals cannot yet be predicted.  Experiments would be particularly revealing for this novel form of matter.    

One can also extend the range of building blocks for liquid-crystal materials.  For example, one could enlarge the particle-size range by using anisotropic colloids.  This leads to a complex space of possible outcomes because aspects, such as shape, flexibility and chirality, of the individual particles matter.  These shapes can lead to new possibilities for self-organization.  

The development of large-scale atomistic molecular simulations could facilitate molecular design and the material synthesis necessary to create materials with novel characteristics.  For example, shape-changing liquid crystals would provide new mechanisms for collective reorganization; by altering the structure one could induce a change of phase in the material.  Another option for triggering large-scale reorganization from small changes in input, is to use lyotropic  molecules (molecules that create liquid-crystal phases when in contact with a solvent).  

In liquid crystals, surfaces often determine the orientation of the director field in its vicinity.  Not surprisingly, therefore, interfaces play a central role in determining the properties of a liquid crystal.  An important area for future research is clearly to control surface interactions so that the material will respond in a desired manner to an external perturbation.  New probes of surfaces and greater control over the chemistry at different scales is therefore desired for novel capabilities and applications.  

In order to advance these goals, it is necessary to develop new tools and instrumentation.  One important effort would be to create advanced imaging capabilities so that information can be obtained over a wide range of length scales \textendash~ from the microscopic level of the individual molecules to the macroscopic level of the order parameter fields.  Another challenge would be the synthesis of active liquid crystals.  Here one would need to develop a general, robust and efficient way to drive materials out of equilibrium.

\section{\label{sec:level1}Polymers and Rheology\protect\\}

Polymers are high-molecular-weight molecules that are concatenated sequences of monomers units that may or may not be identical.  They are the basis of synthetic plastics and natural biopolymers.  These long-chain molecules, with a very high aspect ratio of length to diameter, are difficult to crystallize both due to  steric impurities in the repeat units and to their entanglements; materials made from them are typically disordered and have unique properties due to their length and often-flexible structure.  Of course, there is a large variation in the choice of the component monomers with a rich and flexible set of interactions and novel functional behavior.  Block copolymers, which are polymers constructed of long blocks of two distinct monomer units, can introduce structure on a tens of nanometer size scale.  Polyelectrolytes in solvents such as water can lead to charged polymers.  For some reviews on polymers see chapters in \cite{Cates2000,Kleman2003,Witten2004}.

It has been often quoted that ``There is a great future in plastics.''  Given that the entire plastics industry relies on understanding the properties of polymers, much is known about certain classes of these molecules.  However, given the wide variety of polymeric matter and interactions between polymers and with solvents, there are many outstanding issues.

There are non-trivial implications of topology and architecture for creating new materials.  Star polymers have three or more arms radiating from a single node.  There is an emerging focus on stars in which different arms are chemically different from one another (miktoarms) and complex brush conformations (including bottlebrush).  Such architectures change the thermodynamics of the melts and have implications for the self-assembly of block copolymers and polyelectrolytes where charge positioning, particularly important in batteries, is carefully controlled.  Here, formulating frameworks to describe thermodynamics of charge-containing systems is particularly important.  Liquid-crystal elastomers are materials that merge the orientational order of a liquid crystal into a cross-linked polymer network.  These materials can display extremely large, yet reversible, changes of dimension which is desirable in applications that require actuation \cite{Warner2003}.

Supramolecular polymers, in which the monomers are held together by non-covalent bonds, present further opportunities to formulate new materials because they can undergo reversible transitions between monomer and polymer configurations.  Likewise, in vitrimers polymer networks can be made, broken and reformed due to dynamic covalent bonding.  Applications range from recycling to reinforced tires. Understanding supramolecular properties raises problems in statistical mechanics.  These include how to design charge distributions on brushes and create stimuli-responsive brushes for catalysis. 

Many material properties are determined by disorder including defects in networks, polydispersity in molecular weight and heterogeneity in charge distributions.  Fluctuations influence material  fabrication and properties.  They appear as phase transitions are approached, in the wettability of a surface, in the persistence length of chains, in structural color, and potentially in the permeability of membranes.  Understanding disorder is thus a continuing challenge.

A problem common throughout much of soft matter is to develop a better understanding of non-linear dynamics.  Understanding non-linear interactions from a macroscopic perspective is particularly important in entangled polymers where new models and measurements are required.  This has practical importance for processing polymeric materials.  When materials do not behave as expected for simple flow fields, such as shear or extensional flow, it is particularly important to understand their behavior in mixed fields.  This is especially relevant to manufacturing.  In the case of additive manufacturing, cooperative effects related to rheology of filled polymers, where flow is required at high volume fractions, become important. 

Non-linear dynamics also plays a distinct role in the complex rheology of active materials.  Understanding them fully could lead to opportunities in reconfigurable materials.  In addition, a material can be trained so that it retains a memory of how it was processed.  Glasses, including glassy polymers, display memory in their relaxation behavior.  One intriguing question is whether such memory formation can be used for practical purposes.

As emphasized elsewhere in this report, interfaces are crucial for determining the elastic properties of soft matter.  This is also true for polymeric materials.  For example, interfaces are important for charge transport in electronic polymers, where the polymer/electrode interface must be better understood in the presence of disorder.  An important goal is to create techniques that would allow interfacial processes to be visualized and characterized.  This would yield insight into problems such as stick-slip and adhesion.  One overarching question is to determine how much structural or chemical control is needed over the polymeric building blocks in order to create collective behaviors of interest.  It is unclear whether the interfacial chemistry is sufficient to enable this control.

Synthetic polymers have a very broad range of mechanical properties and are easy to process.  However they are difficult to prepare so as to be absolutely monodisperse.  Building function into synthetic polymers is achieved by adding chemical functionality to a monomer or appending a side chain with a specific functionality. Most synthetic polymers are soluble in organic solvents, although some, like polyethylene oxide, are soluble in water.  On the other hand, biological polymers, like peptides, are truly monodisperse, and assemble into configurations where the three-dimensional arrangement of the chains imparts a specific functionality to the molecule.  They often bind and must be dispersed in water to maintain their well-defined shape. 

There is much to be gained if the synthetic and biological worlds can be merged.   The synthetic polymers, with their ease of processing, could serve to protect the biological polymer in environments other than water.  One could envision synthetic-biological micellar structures in organic or aqueous media where drugs could be hidden within the organic core for delivery.  Conversely, a biological construct could be stored within a polymeric matrix for subsequent use.  One can imagine implants where the structural integrity of a polymer could be coupled with the biocompatibility of peptides or proteins.  There are still few examples where the merging of these two worlds has been effectively accomplished; the potential is great.

To end this section with rheology in its title, mention should be made of shear thickening and shear thinning suspensions.  It is still debated to what extent shear thickening is due to hydro-clusters \cite{Cheng2011} and to what extent it is a jamming phenomenon that requires an understanding of the role of surfaces \cite{Brown2012}.  Magneto- and electro-rheological fluids also have rheological properties that can be varied.  These suspensions contain polarizable particles that respond to the application of an external field.  For example, the application of a magnetic field can alter the rheology of  a magneto-rheological fluid  by changing its viscosity \cite{Wereley2013}.  It has been pointed out that there are intricate interactions between the fluid interfaces, the suspension particles, and the applied field \cite{Orellana2013}.  

\section{\label{sec:level1}Fluids\protect\\}

The subject of fluids has a long and rich history that has impacted and drawn from many disciplines.  Historically, fluid dynamics was one of the frontiers where the mechanics of continuous media was unraveled and continues to be a cornerstone for various branches of chemistry and physics.  Fluid flows are essential for creating many mundane marvels of our daily existence and for producing phenomena observed in more esoteric environments encountered in geophysics, astrophysics and engineering.     

There is an important overlap between researchers in fluids and those in soft matter.  Fluids are involved in many aspects of soft matter: colloids, foams, liquid crystals, polymers, self-assembly, active matter, biomaterials, and microfluidics.  Perhaps more significantly, as emphasized in the introduction, many scientific issues that are important for understanding fluids have resonance within the soft-matter community.  For example, one needs only to supercool a liquid to see a strong increase in its viscosity and a transition into a glass phase.  One can also drive a flowing liquid so hard that it becomes turbulent and new structures, such as eddies and vortex lines, form.  In each case, the liquid is in a state that is far from equilibrium and displays non-linear response. Even after decades of intense effort, our understanding of the glass transition and of turbulent flow are unsatisfactory.

\begin{figure}
\includegraphics[width=0.95\columnwidth]{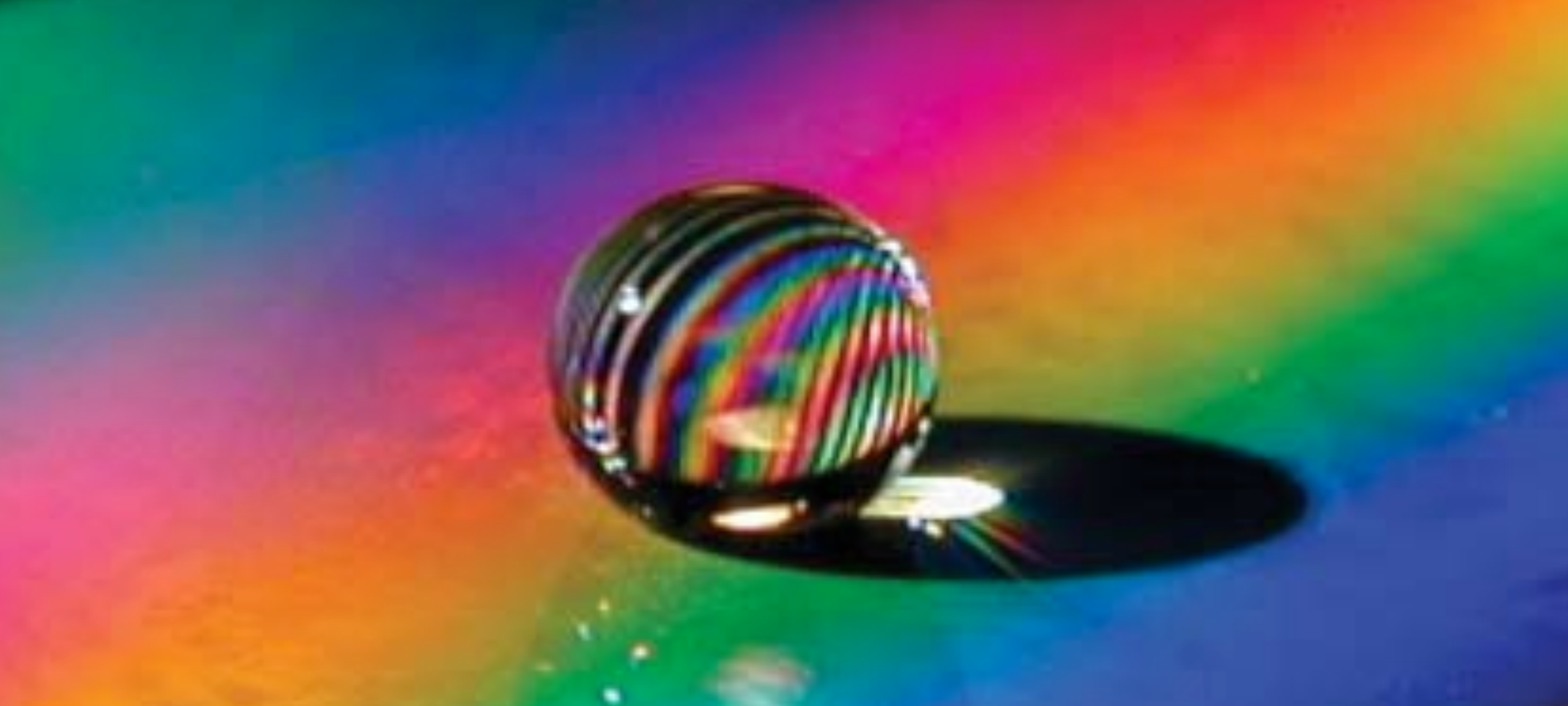}
\caption[]{\label{fig: Figure Droponsuperehydrophobicsurface}
Millimeter-size water drop on a hydrophobic surface textured with regularly-shaped micropillars.  The regular spacing of the pillars creates colored stripes.  From \cite{Callies2005}}
\end{figure}

A rich set of questions await an understanding in fluids that are forced into complex situations as can occur at a fluid interface.  Examples occur when a fluid interacts with a surface (such as a super-hydrophobic surface \cite{Callies2005}) shown in Fig.~\ref{fig: Figure Droponsuperehydrophobicsurface} 
or when several fluids coexist at a line (such as at contact line on a surface \cite{Guo2013,Kaz2012,Style2013}).  Free interfaces present the added complication that they can evolve with the internal motion of the liquid as shown in Fig. ~\ref{fig: Figure_SoapflagZhang} and thus provide dynamically evolving, nonlinear, geometric constraints on the flow.  This creates a highly coupled problem that can often lead to effects such as the formation of surface singularities where two liquid drops coalesces or one drop breaks apart \cite{Shi1994,Eggers1997}.  Some of this singular behavior has been worked out in recent years but certainly much remains to be done especially with non-Newtonian fluids.

\textit{Beyond Newton}:  Much previous research has dealt with simple ``Newtonian'' fluids (those that can be adequately described by a viscosity, a density and a surface tension).  The future of fluid dynamics in soft matter calls for going beyond this restriction and understanding non-Newtonian fluids and the effects that complex molecules can exert on the bulk and surface properties of a liquid.  

\begin{figure}
\includegraphics[width=0.95\columnwidth]{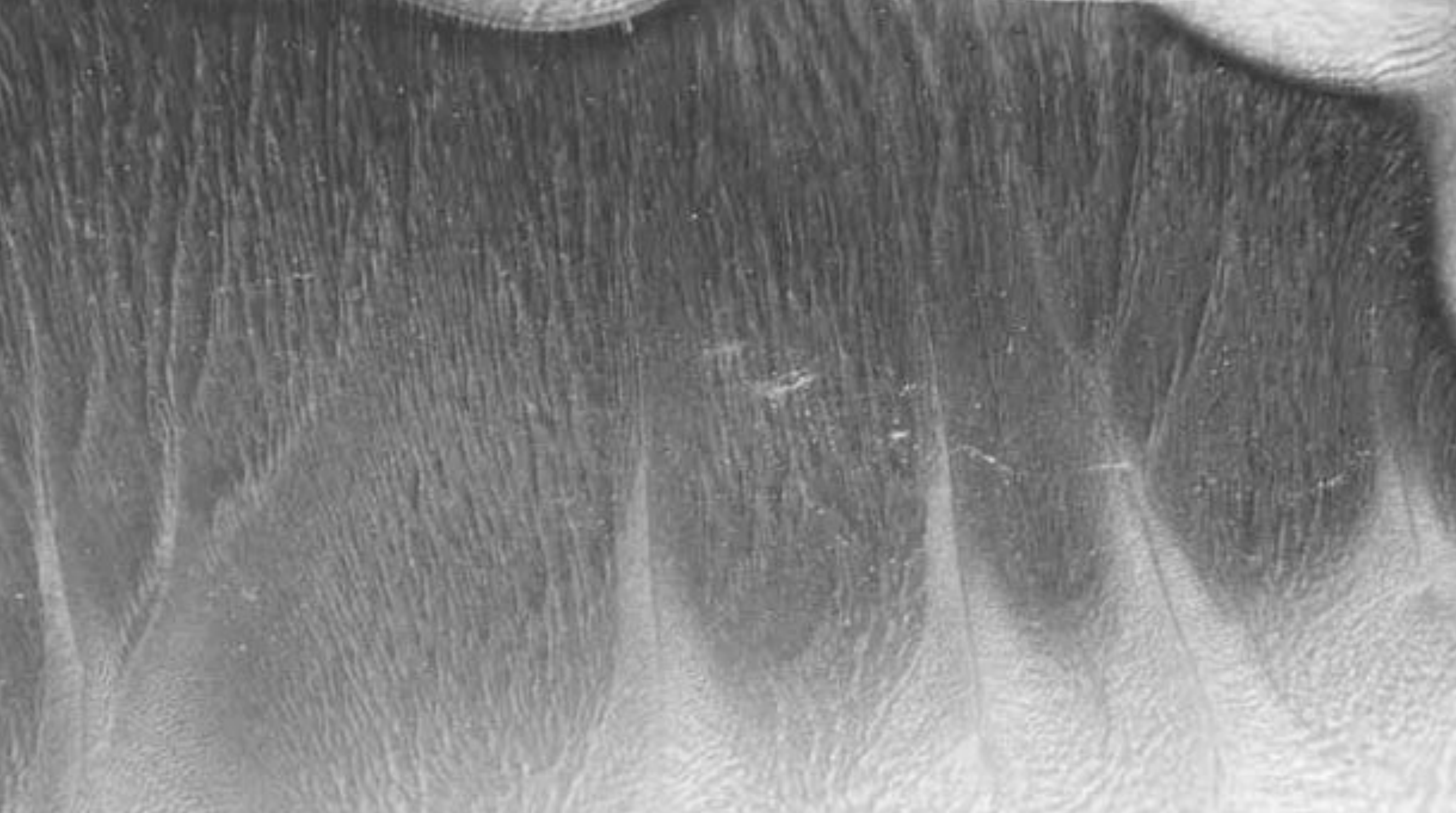}
\caption[]{\label{fig: Figure Tearsofwine}
Tears of wine.  The image shows the tear line is at the top, a thin film region in the middle, and four ridges at the bottom right.  From \cite{Hosoi2001}.}
\end{figure}

It is well known that a variation of the interfacial tension will set up complex flows along a surface.  These Marangoni flows can lead to many counter-intuitive phenomena such as the tears of wine often observed at dinner parties.  There, above the wine's surface, a thin band of wine (composed of alcohol and water, two liquids with different surface tensions) drains down the side of the glass in the form of tears.  An example is shown in Fig.~\ref{fig: Figure Tearsofwine} \cite{Hosoi2001}. The problem becomes compounded when large surfactant molecules are introduced at a fluid interface.  The surfactants alter the local surface tension creating fluid motion, which in turn causes the surfactants to flow.  Research in this area continues to be vibrant.  The coupled problem of surfactants interacting with the flows in the liquid leaves a great deal to be understood especially when the boundary geometry of the medium is involved.  

\begin{figure}
\includegraphics[width=0.95\columnwidth]{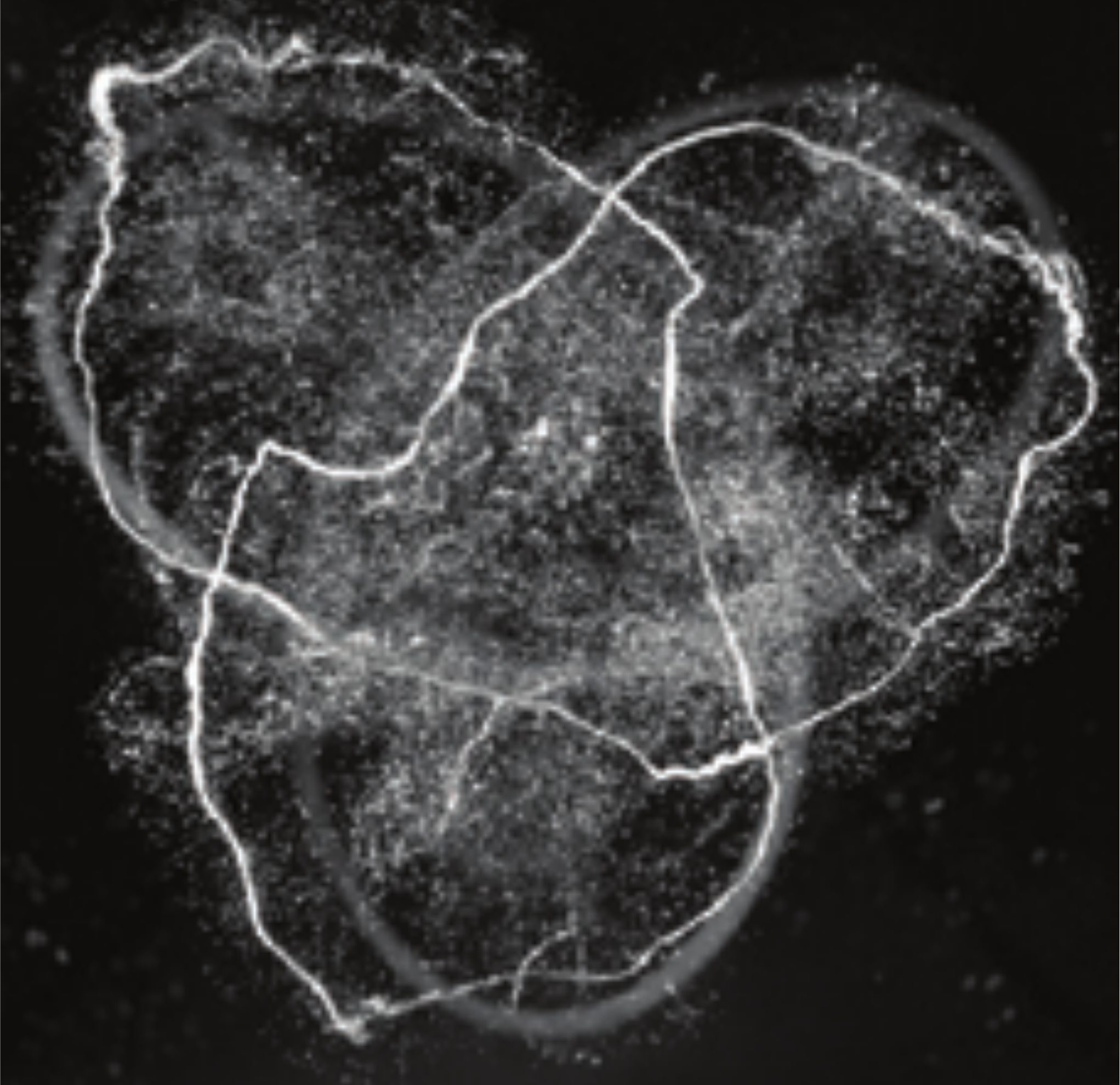}
\caption[]{\label{fig: Figure Vortexknot}
Vortex loop in water created by an accelerated airfoil with the topology of a knot.  From \cite{kleckner2013}.
}
\end{figure}

Adding large active molecules into the bulk of a liquid, as shown in Fig.~\ref{fig: Figure ActiveLC}B, creates internally driven motion in which topological defects in the order parameter interact directly with the flows \cite{Sanchez2012}.  Another example of topological considerations interacting with fluid dynamics is in the case of vortex formation.  Much as magnetic field lines, vortex lines either extend to the edge of the system or wrap around until they join back on themselves.  This is similar to the complex knotting created out of defect lines in liquid crystals mentioned previously \cite{Tkalec2011}.  Such vortex lines can become highly entangled and it can be asked whether a fluid can evolve so that the vortex lines reconnect with themselves allowing the tangle to relax to a simpler geometry.  Studies, as shown in Fig.~\ref{fig: Figure Vortexknot}, show how vortex rings can be tied into topological knotted structures that can subsequently relax \cite{kleckner2013}. 

There are many industries that design fluids with orthogonal required functionalities (\textit{e.g.}, drilling mud, conditioner, or toothpaste) but these are fluids are typically designed empirically: ``add things until it works''.  There are currently no universal design principles to achieve multiple functionalities in a single fluid.  The creation of such an overarching set of rules is one outstanding challenge for the field.  

\section{\label{sec:level1}Mechanical Meta-materials\protect\\}

A rapidly growing research area deals with the design and fabrication of composite materials that, due to their assembled structure, have unusual elastic properties.  In these materials, the behavior and response depend on how the many components are assembled rather than on the properties of the individual components themselves.  The advantage of such a design strategy is that one can attempt to create materials with a set of desired properties in mind that may be difficult to find in nature.  An example that has become popular is to create materials with negative Poisson ratios (\textit{i.e.}, auxetic response in which compressing a material along one direction decreases the other two dimensions as well) \cite{Greaves2011}.  While negative Poisson ratio materials do exist naturally, they are less common than the normal approximately incompressible ones.  It is an open question about how much freedom one has to design in multiple distinct properties into a material.

The disadvantage of meta-material design is that one has to develop a means to guide the assembly of pieces into the determined structure.  When the individual pieces themselves are microscopic, then this can be a challenge.  However, if the pieces can be coaxed to self assemble in the proper manner, or if only their shape but not their precise placement is important, this difficulty could be circumvented and a significant problem would be solved.  With the advent of 3D printing technology many different structures can be built Ð at least at a moderately small scale.  One challenge is to push such technology to create much larger structures with ever finer features.  Even if that were to become possible, it can still be asked whether such technology could mass-produce material in sufficient quantity to be technologically useful.  

A variety of routes to meta-material design are currently being exploited to control the shape, mechanical properties or actuation of a material.  One of these is based on the paradigm of origami that models three-dimensional structures by folding a two-dimensional sheet.  (If the sheet has prescribed cuts this is called kirigami.)  Because a fold along one crease constrains the ability of the sheet to distort elsewhere, an open theoretical challenge is to develop the rules for folding that do, or do not, produce bending of the flat sections of the sheet.  It is important to understand what are the degrees of freedom accessible to a folded sheet \cite{Silverberg2015}.   From the experimental perspective with which this report is predominantly concerned, one wishes to create such objects efficiently from the designs that have been evolved theoretically or computationally.  More exciting would be to find ways in which the material itself can develop the desired folds to allow the emergence of unusual and desired behavior.

Using evolutionary computer algorithms, one can simulate what shapes of particles can lead to a desired rheological behavior. One wants to discover a microscopic building block that will achieve a desired outcome.  This has been done in designing a shear-thickening granular material with an unusual stress-strain curve in which the material gets progressively stiffer as the material is sheared \cite{Miskin2013}.  How far can such ideas be pushed to create novel and useful materials \cite{Ohern2013}? 

Another tactic has been to design properties in network structures.  Pruning specific bonds in computer-simulated networks derived from jammed packings allows the Poisson ratio  to be varied at will \cite{Goodrich2015}.  Again, from an experimental point of view, the issue is how the material can be coaxed into pruning the appropriate bonds itself without resorting to a simulation.  Such results raise the issue about what are the limits to which function can be programmed into a material.  The Poisson ratio is a global property. Can similar control be obtained at a local level, for example, as exemplified by biological proteins with allosteric behavior?  There, binding a molecule to one specific site determines whether a far-distant site will have the ability to bind to a third molecule.  Can similar allosteric responses be designed into physical materials?  The answer appears to be in the affirmative \cite{Rocks2016}.  Likewise material failure Ð how cracks propagate when a material is stressed to the point of failure Ð also appears to be related to the average number of constraints in the material \cite{Driscoll2016}.  Topological meta-materials also require understanding how the degrees of freedom (allowed particle motions) interplay with the inter-particle constraints \cite{Kane2014,Paulose2015,Nash2015}. These examples, based on understanding the constraints in a network, may be just the opening wedge of what kinds of functions can be imparted to a material by careful manipulation of the bonds.  

\begin{figure}
\includegraphics[width=0.95\columnwidth]{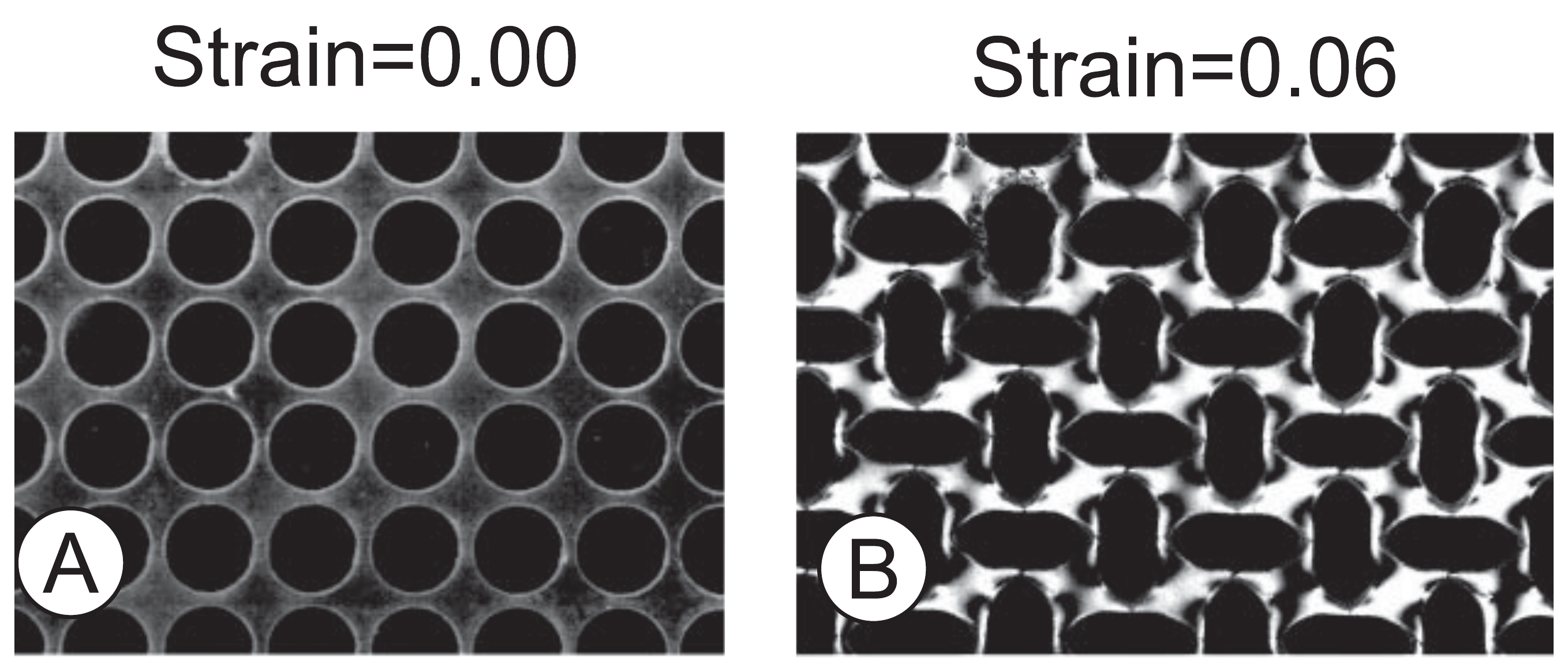}
\caption[]{\label{fig: Figure Auxeticfoam}
(A) Thick sheet with circular holes.  (B) This geometry produces a negative Poisson ratio well into the non-linear regime; as the material is compressed in one direction, it is compressed in the perpendicular direction as well.  The material is birefringent under strain so that the stresses can be visualized.  From \cite{Mullin2007}. }
\end{figure}

The field of meta-materials is still in its infancy and it is impossible to tell with certainty where research will lead.  It is possible to design materials with a desired property either in its linear or in its non-linear response.  Both behaviors could be useful in different contexts.  Going beyond linearity poses an overarching challenge.  The material often undergoes catastrophic failure once the linear regime has been exceeded.  Yet one would often like to avoid failure but still make use of large-scale response.  A beautiful example of a material that has a large non-linear auxetic response is shown in Fig.~\ref{fig: Figure Auxeticfoam}.  A square lattice of holes drilled into block will break the symmetry under uniaxial compression and produce a negative Poisson ratio for large amplitude deformations \cite{Mullin2007,Florijn2014}. 

\begin{figure}
\includegraphics[width=0.95\columnwidth]{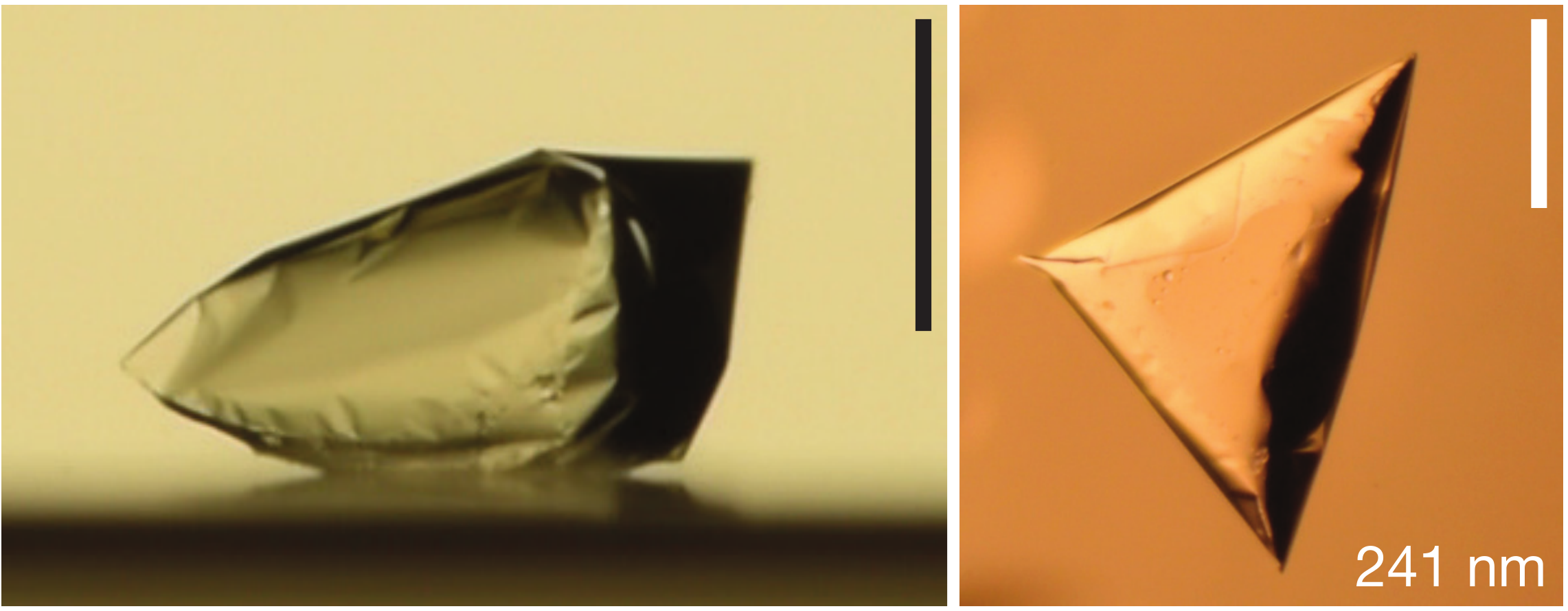}
\caption[]{\label{fig: Figure Wrappingliquids}
Side and top views of a circular polystyrene sheet wrapping a water drop immersed in silicone oil.  Sheet thickness is $241$ nm.  Scale bar is $1$ mm.  From \cite{Paulsen2015}.}
\end{figure}

Another goal is to create materials that can transform their shape and function in response to a stimulus.  One example is a self-folding surface such as occurs when an ultra-thin sheet encapsulates a liquid drop as shown in Fig.~\ref{fig: Figure Wrappingliquids}. This large-scale motion takes advantage of the ability of thin sheets to have large distortions without an enormous energy penalty \cite{Paulsen2015}.

By playing one set of forces off against another, one can add functionality to a material .  When they are in balance small shifts in steady-state conditions can lead to one set of forces overcoming the others.  If one member buckles it can set off an avalanche of activity.   One possibility is to pit surface tension against elastic forces at an interface.  Working near a phase transition is another way that a small change in a control parameter (\textit{e.g.}, pressure, temperature, shear stress) can cause a colossal change in a material.  One can design inhomogeneity into a material to provide  local control.  

Taking lessons from biology, such as in the case of allostery mentioned above, can lead to new forms of behavior.  Basic understanding of how a cell moves or muscles contract can provide ideas for how similar mechanisms can be employed in physical systems.   On an even grander scale, one can conceive of active matter forming the entities that comprise a meta-material. 

This has been just a taste of what approaches can be used to add functionality to a material by assembling and manipulating the elements at a mesoscopic level.   The possibilities seem limitless.

\section{\label{sec:level1}Active and Adaptive Matter\protect\\}

The study of active matter seeks to attain a fundamental understanding of collective properties that emerge in an ensemble of particles or agents each of which consumes energy to propel itself.  Motion arises naturally in such systems and their understanding requires a broad perspective that unifies the global features of the flow with the microscopic propulsive mechanisms responsible for the activity \cite{Marchetti2013}.  We have referred to them throughout this report in the sections on colloids, polymers, liquid crystals, fluids, mechanical meta-materials, self-assembly, pattern formation and granular materials.  The challenge is to determine the fundamental difference between driven non-equilibrium systems, such as turbulent flows, where energy is typically injected at the macroscopic system size and then is dissipated at a smaller level, and active matter where the energy is injected at the microscopic scale and proceeds to generate motion over large distances.  (Figure~\ref{fig: Figure ActiveLC}A  shows a case that perhaps lies between these two extremes.  The entire apparatus is vibrated, but the individual particles couple locally to this overall input of energy in different ways so that they seemingly have their own energy source.)   Active matter is inherently far from equilibrium in an unexplored and interesting fashion so that much of our intuition built upon experience with inactive systems is bound to be inadequate.  This is a new frontier between different subfields of soft matter.  

\begin{figure}
\includegraphics[width=0.95\columnwidth]{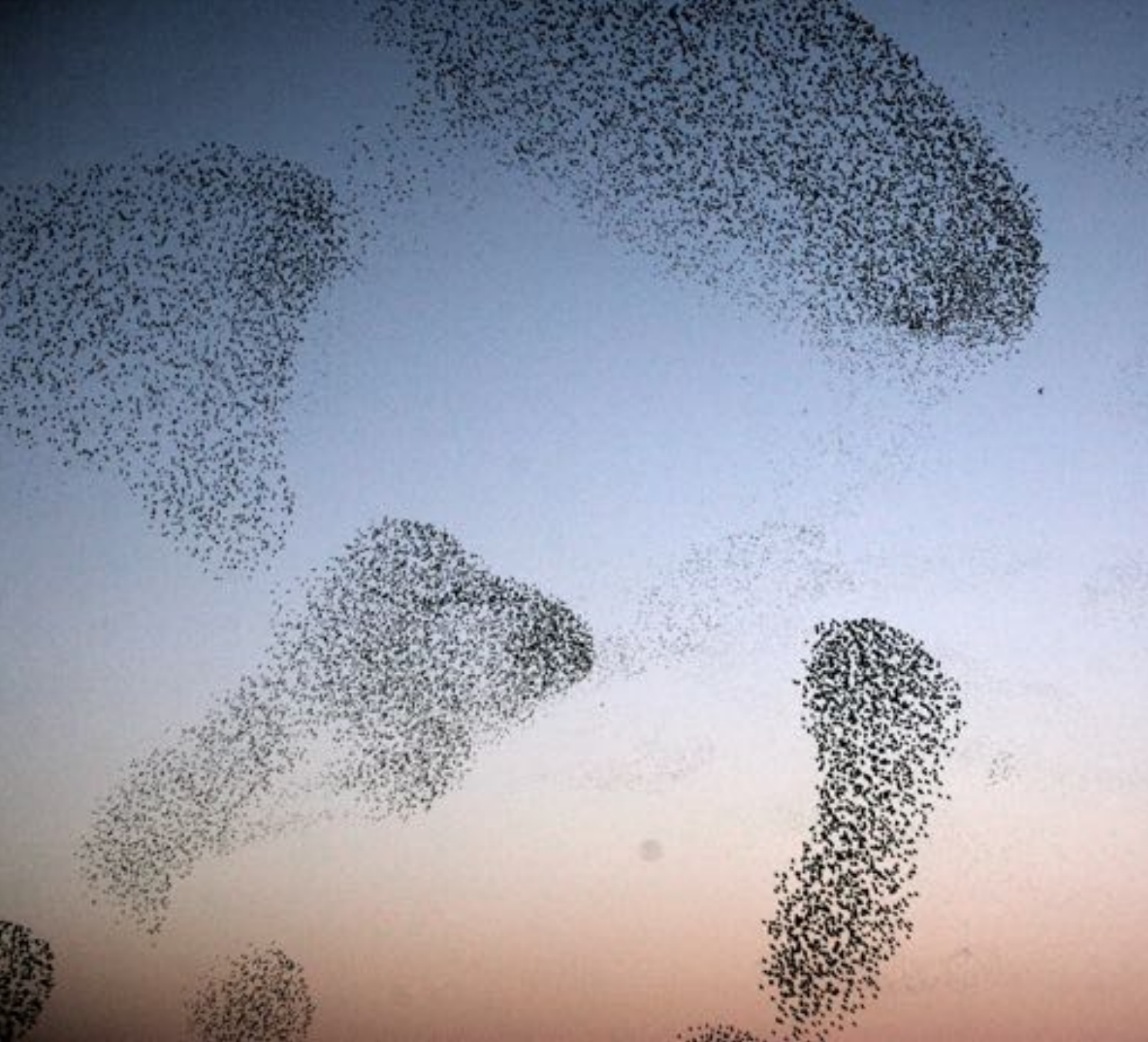}
\caption[]{\label{fig: Figure_Flocking}
Bird flocks over the Rome train station.  From \cite{Cavagna2010}.}
\end{figure}

The most obvious examples of active matter come from the realm of biology.  On the molecular scale, self-organization occurs in actin filaments or microtubules and controls the structure of cells; on a much larger scale, the agents are the animals themselves and their activity leads to the stunning and mesmerizing flocks of starlings as shown in Fig.~\ref{fig: Figure_Flocking} or in schools of fish.  The flocking represents a non-equilibrium transition into a phase that has coexistence of high- and low-density regions and correlated motion of the agents \cite{Cavagna2014,Marchetti2013}.   Introducing active agents generates a new degree of complexity and produces new features not observed in matter based on more conventional inter-particle interactions.  

\begin{figure}
\includegraphics[width=0.95\columnwidth]{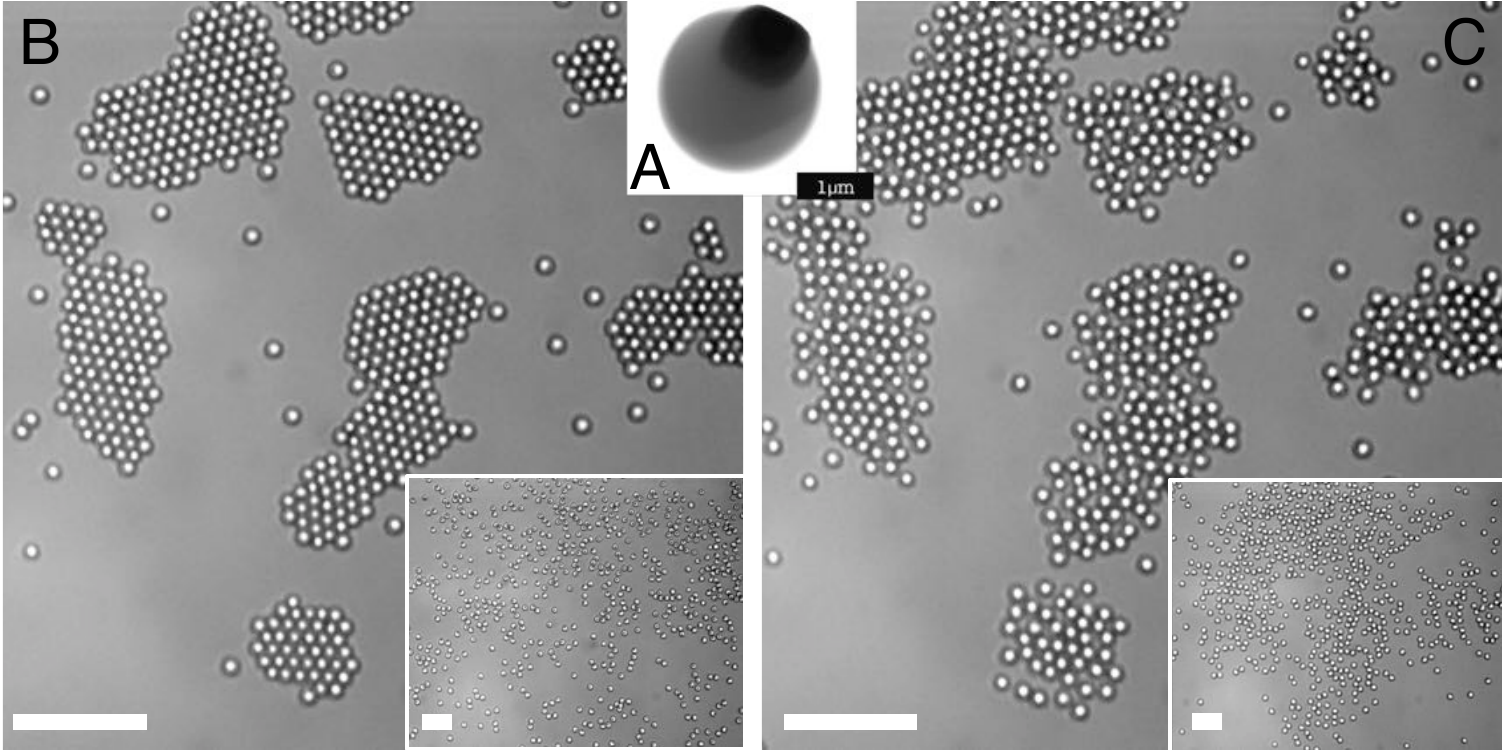}
\caption[]{\label{fig: Figure_Livingcolloids}
(A) Scanning electron microscope image of a bimaterial colloid with a protruding hematite cube.  (B) Formation of a living crystal.  When illuminated, the particles assemble into crystals.  Inset shows initial homogeneous distribution.  (C) When the light is turned off, the crystals melt. Main panel shows after $10$ seconds; inset shows after $100$ seconds.   From \cite{Palacci2013}.}
\end{figure}

Active liquid crystals represent a synergy between different soft-matter efforts including fluid dynamics, rheology, liquid-crystals, polymer physics, biophysics and extreme mechanics including the study of fracture.  One beautiful example of an active liquid crystal is shown in Fig.~\ref{fig: Figure ActiveLC}.  Here the active liquid crystals show buckling, folding, internal fracture of a nematic domain and a pair of topological defects. Another example of self-organization in a synthetic colloidal suspension is shown in Fig.~\ref{fig: Figure_Livingcolloids}B of a two-dimensional ``living crystal'' created from light-activated artificial surfers \cite{Palacci2013}. 

It is interesting to compare the generality of active systems with biological organisms, which are also inherently out of equilibrium.  Biological systems usually employ a limited number of structural and dynamical motifs in order to build diverse functionalities.  These include cell division, motility, and reproduction.  By contrast, active matter goes beyond biology by putting components together in novel ways that had not previously been observed.  It has less constraints.  Studying active matter thus perhaps has the potential to teach us ways of forming structure beyond what biology has done.

One important area is the study of active gels such as occurs when molecular motors using chemical energy drive mechanical motion in a network of cross-linked filaments.  These active gels are important in the dynamics and structure formation in cells.  This is an attractive paradigm for understanding how localized sources of energy can lead to novel and unexpected behavior.  A current challenge is to use the ideas and theories developed in the biological realm to branch into other active materials.  This will require  developing complementary model systems: synthetic active particles, living organisms, reconstituted systems and active granular systems. 

Making new microscopic and/or macroscopic units, which transduce energy, remains a challenge.  It will require establishing new connections between chemists who can make new motors and biologists who can isolate, purify and modify biological force-producing units.  New characterization techniques will also be needed.  As highlighted in the section below on instrumentation, new rheological capabilities would be an important advance. 

\section{\label{sec:level1}Self Assembly\protect\\}

Providing functionality to materials is an outstanding challenge that incorporates the chemistry of the components and the structure of their assembly.  Thus, a common goal of many areas of material science is to assemble matter into useful structures.  It is particularly attractive if the sub-units of the material (atoms, molecules, colloids, \textit{etc.}) were to self assemble into a desired arrangement with as little external control exerted as possible.  It has long been debated as to what constitutes self-assembly since there is always some degree of external control exerted in setting up an experiment.  One seems to know self-assembly when one sees it.  

An atomic crystal may be said to self-assemble as its constitutive atoms arrange themselves into the lowest free-energy state.  But, even there, considerable care and expense must be lavished to create pure, nearly-perfect, crystals of \textit{e.g.}, silicon.  Soft-matter can also self-organize into ordered arrays, albeit not as perfect as the crystals used in the semiconducting industry, by minimizing their free energy; $k_{B}T$ is important.  Other structures form spontaneously.  For example, micelles self-assemble in colloidal suspensions of surfactant molecules.  Other structures such as vesicles or membranes can also form \cite{Witten2004}. 

One set of open questions in self-assembly now concerns the rules of assembly when $k_{B}T$ is  \textit{not} important \cite{Cheng2006,Hu2014}.  An increasingly common route to fabricating desired structures is to use dynamics as a tool to force the components to assemble in certain directions or with certain patterns.  There are a variety of dynamical variables that can be controlled.  In addition to temperature, these include vibration (for large particles such as in granular media), shear, compression, external fields, optical tweezers, and using active particles (including robots for larger entities Ð for example on an architectural scale \cite{Augugliaro2014}).  Particle shape is clearly an important parameter that controls the aggregate structure  \cite{Glotzer2007}.  A better language than we currently have, including a better understanding of the statistical mechanics and thermodynamics for these far-from-equilibrium, driven, or glassy systems, is necessary to describe such evolution.   For example, one would like to generalize the concept of minimizing the free energy to bring the collective system to a specific state.  

Bringing ideas and tools from biology, where assembly of specific materials and functions is constantly occurring, into soft matter is a continuing goal.  Likewise, programmable self-assembly would allow the construction of more nuanced materials.  For example, large two-dimensional lattices have been constructed by manipulating the interactions at different length scales by using biological molecules such as DNA \cite{Reif2001,Seeman2003,Wang2015}.  

\section{\label{sec:level1}Patterns and Structure Formation\protect\\}

Patterns emerge from an apparently featureless background.  They are everywhere and many have their provenance in soft matter.  They play an important role in many subject areas outlined in this report: liquid crystals, active matter (see Fig.~\ref{fig: Figure ActiveLC}), fluid mechanics (see Figs.~\ref{fig: Figure_SoapflagZhang}, ~\ref{fig: Figure Tearsofwine} and ~\ref{fig: Figure Vortexknot}), foams, packings and colloids.  They are the outcome of design using origami and kirigami.  

\begin{figure}
\includegraphics[width=0.95\columnwidth]{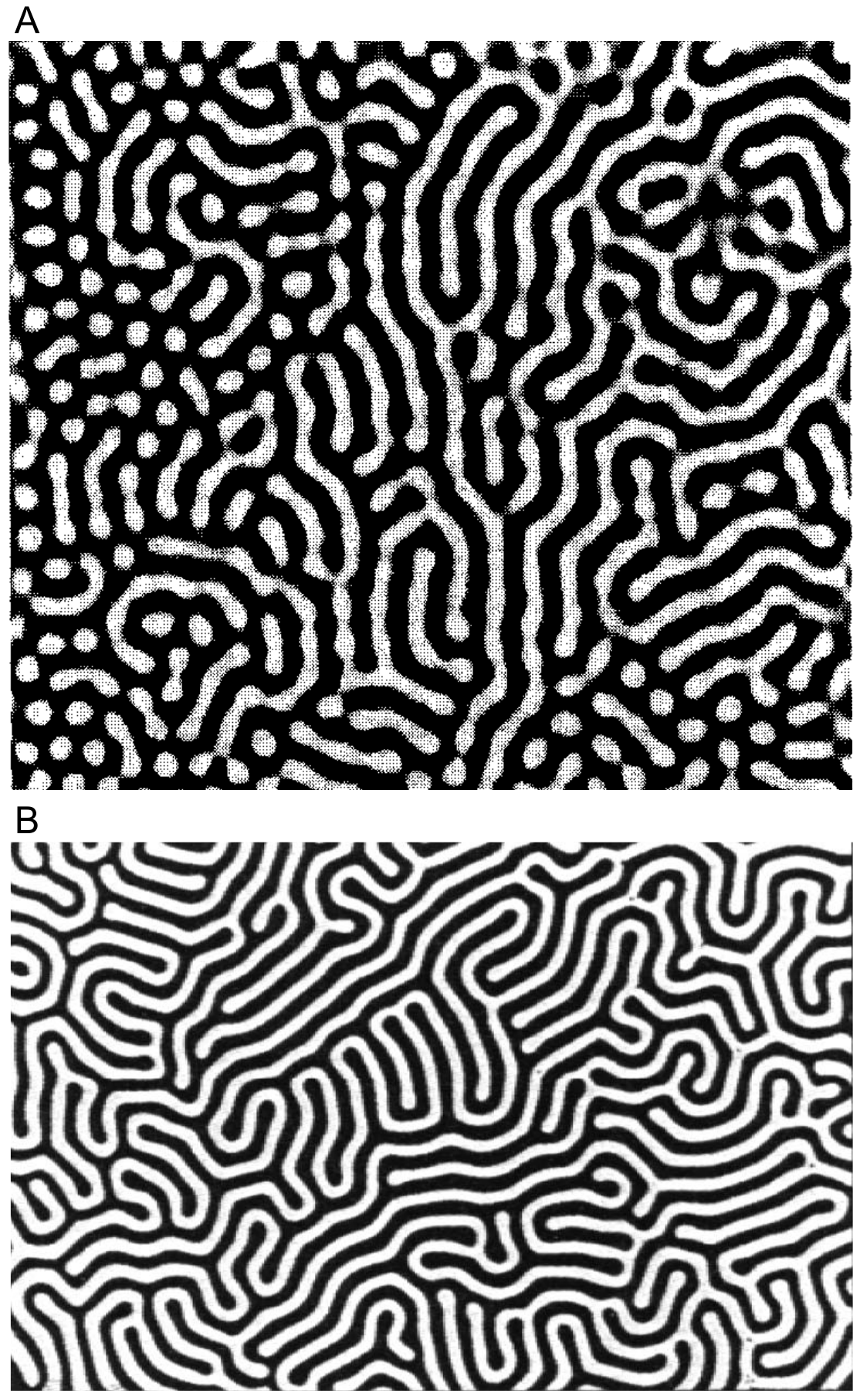}
\caption[]{\label{fig: Figure Labyrinthinepattern}
Labyrinthine patterns.  (A) Turing pattern.  Stationary chemical patterns formed in a gel layer with striped patterns with wavelength $0.52$mm.  From \cite{Vigil1992}. (B)  Modulated phase. A thin layer of ferrofluid in a magnetic field with a period $\approx 2$mm.  Adapted from Fig. 4 of \cite{Rosensweig1983} as in \cite{Andelman2008}.
}
\end{figure}

Certain patterns are familiar and appear recurrently in different guises.  For example, although not identical in all cases, branching appears in river networks, viscous fingering, blood vessels and of course in tree branches and dielectric breakdown, (see Fig.~\ref{fig: Dilationsymmetry}).  Another pervasive example is based on the Turing reaction-diffusion mechanism \cite{Turing1952} which have been demonstrated in chemical reactions \cite{Ouyang1991}.  An example is shown in Fig.~\ref{fig: Figure Labyrinthinepattern}A.  It has been suggested that such Turing patterns are responsible for many biological shapes and forms such as pigmentation in the coats of animals (for example zebras or leopards) and patterns on sea shells.  Similar looking labyrinthine patterns are seen in modulated phases \cite{Andelman2008}.  One example is in thin layers of ferrofluid (suspensions of a magnetic solid in a fluid solvent) in a magnetic field shown in Fig.~\ref{fig: Figure Labyrinthinepattern}B 

In some cases the initiation of patterns is caused by an instability as some control parameter is varied past a threshold value.  When there is a most unstable wavelength at the threshold, a single length scale characterizes the onset of patterns throughout the system.  This occurs, for example, in viscous fingering where the finger width is set by a competition between stabilizing surface tension and destabilizing viscous stresses \cite{Saffman1958}.  A related example occurs when the non-linear dynamics of the system are controlled by the presence of a nearby singularity such as occurs when one liquid drop fissions into two or more smaller pieces \cite{Shi1994,Eggers1997}.  In that case, the Rayleigh-Plateau instability sets the scale for subsequent droplet formation.  In other cases, such as the diffusion-limited aggregation in Fig.~\ref{fig: DLA}, the structure is apparently fractal with no characteristic length \cite{Witten1981}. 

The characterization of a pattern requires determining its symmetries and conservation laws and identifying what order parameter describes its essential behavior.  Sometimes the order is not obvious and new metrics need to be invented to describe them as in the case of hyper-uniform density fluctuations \cite{Torquato2016} or in the wavy edges of a tree leaf \cite{Sharon2007}.  Often the patterns are caused because the system is driven far from equilibrium, as in turbulent flow, so that new structures appear \cite{Zocchi1990}.

Unlike amorphous materials, crystals have well-defined defects \cite{Chaikin2000,Kleman2003}.  These defects, which often have a topological character, control deformation as the material responds to strain and determine the  patterns that can be formed.  This is seen dramatically in lamellar layers of block co-polymer or liquid-crystal films.  The behavior of the defects is intimately tied to the presence and location of surfaces in finite-sized materials.  In liquid crystals, the way in which the director field interacts with the bounding surface of the liquid often requires that the defects must exist in the interior.  Such defects play a special role in the dynamics. 

\begin{figure}
\includegraphics[width=.95\columnwidth]{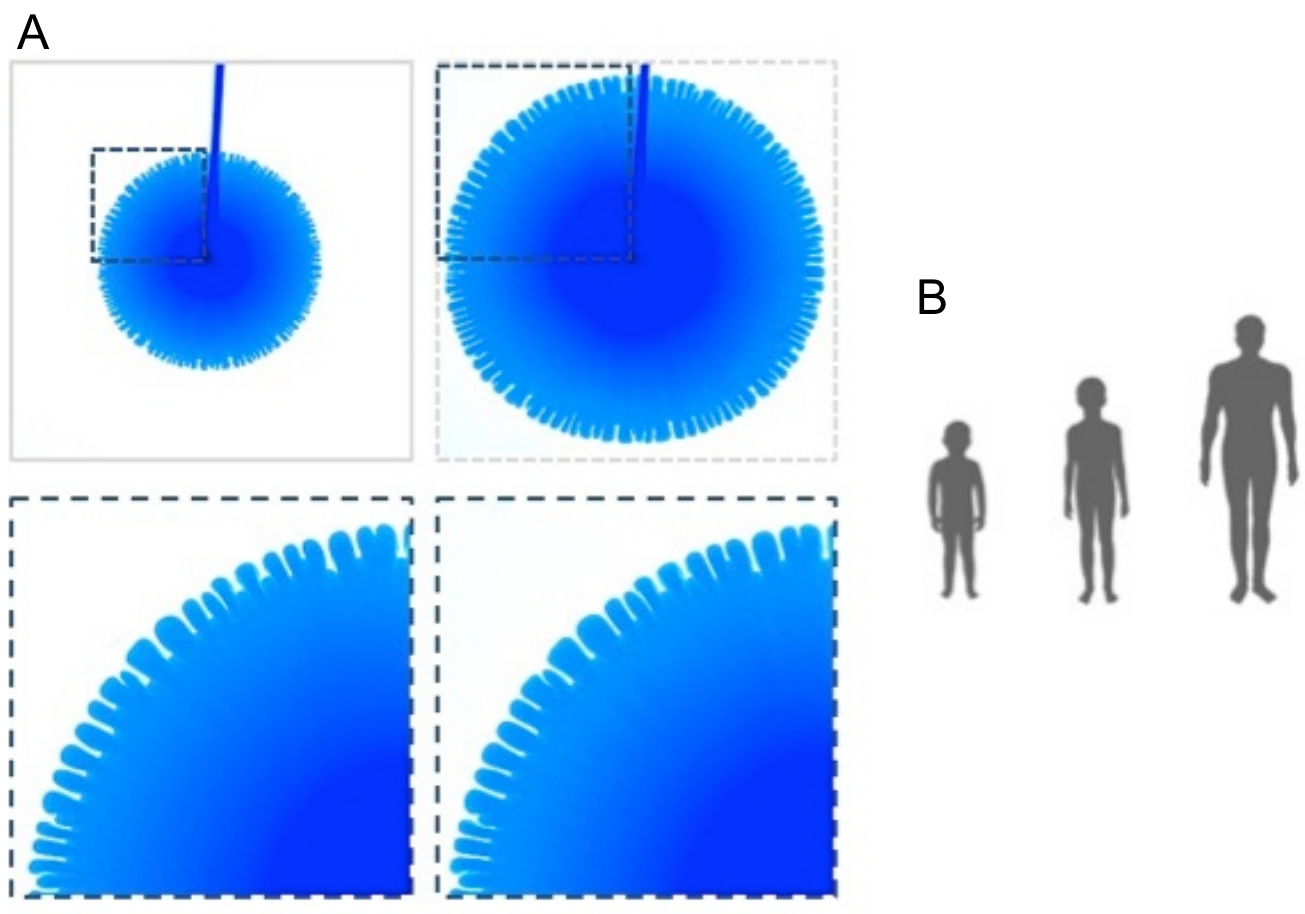}
\caption[]{\label{fig: Figure_Proportionategrowth}
Proportionate growth in miscible viscous fingering.  (A) The blue (darker) fluid displaces a more viscous fluid with which it is miscible.  The interface between the two fluids at an early time can be enlarged so that it lies on top of the interface at a later time.  Even the small features are nearly identical after the images have been scaled.  (B) This form of pattern is common in the biological world (\textit{e.g.}, mammalian growth), but few physical examples exist.  From \cite{Bischofberger2014}.}
\end{figure}

Far from being a closed field, new patterns are frequently being discovered.  The example of proportionate growth shown in Fig.~\ref{fig: Figure_Proportionategrowth} is an example of a pattern that had not been observed in the physical realm prior to recent experiments on miscible viscous fingering \cite{Bischofberger2014}.  Other examples are surely waiting to be discovered.  Likewise, well-known mechanisms to create patterns, such as the reaction-diffusion mechanism of Turing, constantly find new applications in nature.   Indeed biology is a fecund place to search for a variety of structures \textendash~ both new and old.  In order to take advantage of many of these opportunities and to search for new forms of structure, it is important to understand the organizational principles that appear in dissipative systems and that govern non-equilibrium assembly processes.

One avenue for productive exploration is the use of pattern formation as a means of imparting functionality to a material.   Again, one can take lessons from biology.  For example, inspired by embryogenesis, one could attempt to use active growth patterns to localize buckling and folding so as to create and control new structures.  One could build shape-morphing structures such as PDMS gels or ultrathin sheets that are able to encapsulate liquids droplets (see Fig.~\ref{fig: Figure Wrappingliquids}).  

One could also try to create functionality on new length and time scales.  One 
possibility would be to employ agents with feedback that could cycle between different states of the system.  These could provide reprogrammable devices or the ability to create decision-making materials.  

\section{\label{sec:level1}Simulations and Big Data\protect\\}

The sheer amount of data collected in a single soft-matter experiment can be staggering.  For example, many experiments depend on high-resolution, high-speed video.  Manipulating a data file for even one such movie in order to identify features over hundreds of thousands of frames is a daunting challenge.  The era of big data has arrived in soft matter.  

In addition, soft-matter systems are so complex that it is often difficult to extract from experiments the essential roots of their distinctive behavior.  As the understanding of a system grows, a challenge is to devise novel methods that can identify new objects or descriptors that capture the essence of the underlying science.  Significant advances along this line have been made including the use of machine learning, evolutionary algorithms, computational homology and network methods.

Computer simulations have an advantage over experiment in that they can readily investigate a series of model systems by selectively adding or subtracting degrees of complexity in the interactions between particles or by selectively tuning control parameters.  Thus for granular media simulators can vary particle shape, gravity and friction coefficient; for polymers they can vary chain length and cross-linking; for liquid crystals they can vary molecular architecture.  Because soft-matter is classical, one does not have to resort to quantum algorithms to do an accurate simulation.  The sophistication of both computer hardware and modeling techniques continues to increase rapidly so that simulations have an ever-increasing role to play in discovering new soft-matter phenomena.   

Many groups now use simulations as a standard tool rather than as a research focus.  The codes for molecular dynamics simulations are widely distributed and used.  However, as investigators move towards more parallel processing, there is less unanimity.  Some groups use graphic processor units (GPUs) and others use standard computers with \textit{e.g.},, the LAMMPS code \cite{Frantzdale2010}.  In simulating liquids, there are again a variety of techniques to simulate, for example, the Navier-Stokes equations.  These are rapidly getting more powerful with the ability to handle more complex problems with multiple fluids.   

There are, of course, new challenges.  In order for a simulation to be reliable, it must be fully validated.  In order for a simulation to be able to demonstrate novel physical effects, it is essential to know what features must be included at the microscopic level such as inter-particle potentials and at the macroscopic level such as boundary conditions.  The best way to validate a code is to compare the simulations with benchmark experiments.  This must be done on a field-by-field basis; it is valuable to have standardized codes where possible and to identify and agree on canonical experimental systems (boundary conditions, external parameters) and make sure that the data generated is reproducible.  In soft matter, the synergy between experiment and simulation has been excellent; strong collaborations between experimental and simulation groups should continue to grow.  

While some large databases exist, such as JHTDB (Johns Hopkins Turbulence Databases) \cite{Li2008}, the soft-matter community does not generally possess and query large databases that can be used to shorten the time and cost of materials development.  A better infrastructure is needed to create additional databases of soft-matter experimental and simulational N-body systems.  The question then arises of what type of data should be included in such databases.  At one extreme, one could include all the microscopic positions and inter-particle forces; at the other, one could include only the macroscopic outputs such as stress versus strain.  Because the system response can be spatially inhomogeneous, each deformation can be different and one must know the complete history of a system and the spatial dependence of its response. Much can be learned from experience with previous databases. 

\section{\label{sec:level1}Instrumentation\protect\\}

As with any branch of experimental science, significant and rapid progress often depends on the creation and development of new instruments that allow measurements that had previously been impossible.  Soft-matter science has had an excellent history of developing tools that have allowed new and precise methods of probing and manipulating matter at the mesoscopic scale.  These include the invention of diffusion-wave spectroscopy (DWS) \cite{Maret1987,Pine1988} making it possible to detect minute changes in the structural arrangements of a macroscopic colloidal sample on a rapid timescale.   Sophisticated laser-tweezing technology (and related methods that use the forces from a light beam) can now manipulate enormous numbers of particles simultaneously \cite{Grier2003}.  The emergence and widespread use of high-speed confocal microscopes allows precise depth resolution of structure and the tracking of individual particle positions in both soft and biological matter.  Other means of visualization include holographic microscopy (for reconstructing three-dimensional volumes), magnetic resonance imaging (for imaging particles and fluids containing oil or water), X-ray tomography, high-speed video photography, photoelastic visualization (for imaging the stress-induced birefringence of a two-dimensional sample), and light-sheet fluorescence microscopy (for mapping three-dimensional particle positions from two-dimensional slices).  Scattering techniques have also led to important advances; for example, neutron scattering allows the examination of individual polymer chains by deuterium labeling and neutron reflectivity provides sub-nanometer resolution of soft matter at surfaces and interfaces.  Soft X-ray scattering enables a determination of the structure and morphology of complex soft-mater systems with exceptional spatial resolution.  This is not an exhaustive list.

Challenges remain in order to make the next advance in experimental capability.  While one cannot predict where the next breakthrough will occur, we can lay out some of the areas in which progress would be welcome.  

\begin{figure}
\includegraphics[width=0.95\columnwidth]{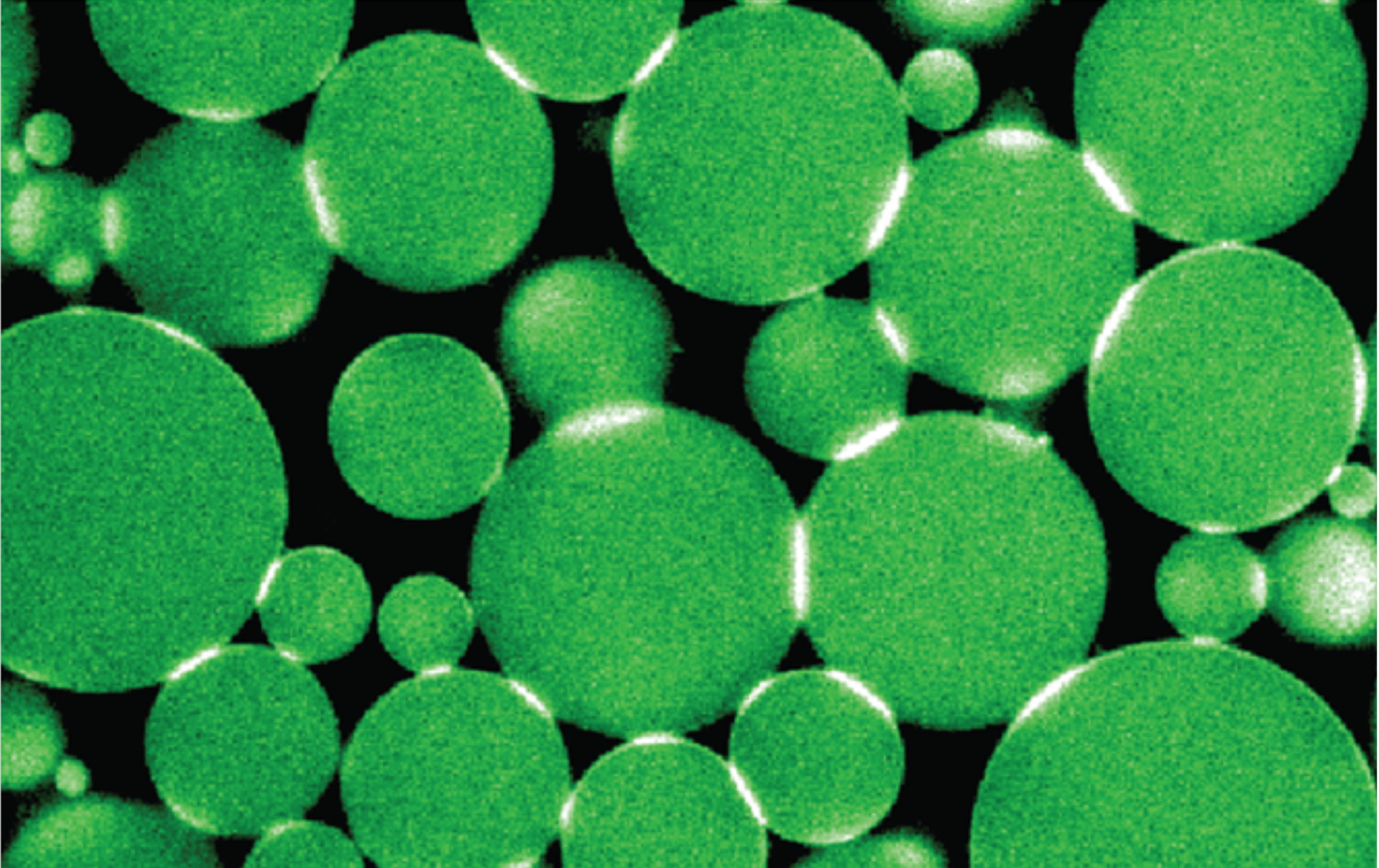}
\caption[]{\label{fig: Figure Colloidalforces}
Confocal imaging of a suspension of droplets.  The bright spots, where the green (lighter colored) droplets contact each other, allows contact forces between the suspension droplets to be measured.  From \cite{Brujic2007}.}
\end{figure}

It is common to image a sample in two dimensions or at the surface of a three-dimensional material.  It is more difficult to resolve simultaneously the dynamics and stresses acting at the particle level in the interior of a sample.  One of the most used tools in colloidal science is the confocal microscope which allows unprecedented spatial resolution.  While commercial machines can now operate at $1000$ frames per second, higher speeds are often necessary to capture fast dynamics.  Figure~\ref{fig: Figure Colloidalforces} shows a confocal-microscope image that led to a measurement of contact forces in a static suspension.  One challenge is to create a technique that can make such measurements rapidly on a moving material in an opaque medium.  

As remarked in the introduction, soft matter is often also slow matter.  Thus to watch a sample equilibrate, or even to detect the transient behavior adequately, it can be necessary to extend experiments over exceedingly long times.  Some spectroscopic measurements have studied temperature-controlled glasses over periods of a month \cite{Leheny1997}; it is prohibitive to extend that time significantly.  The problem is compounded when the range of times is coupled with an equally large range of lengths from $0.1 nm$ to $1 m$.  That is we are asking (!) for techniques that allow an order of magnitude in the order of magnitude of lengths and times.  

The tactic often used in a statistical approach to a many-body system is to concentrate on average quantities.  However, in soft-matter systems, it is often important to resolve the behavior on the individual particle scale as well.  The data sets become enormous when one needs to detect many particles simultaneously over long times.  It is still an art to handle such big data sets and new computers and algorithms are needed to make such analyses routine.  This is not only a problem faced by the soft-matter community, but by other research and development areas as well.   A synergy with the computer-science community has already had large benefits in soft-matter regarding simulation techniques; graphic-processing-unit (GPU) computers, originally developed for a different end use, has made parallel-processing easier and faster.  Such hardware advances are also useful for implementing image-processing algorithms.  Likewise, open-source software has provided solutions to many programming challenges.

As has been emphasized repeatedly throughout this report, the activity of a soft material often involves its interfacial properties.  The motion of contact lines is but one example that has many consequences \cite{Guo2013,Kaz2012,Style2013}. Studying interfaces requires tools that are specifically sensitive to the surface to obtain information about the local structure, local rheology and local excitations.  Being able to image and characterize buried interfaces is a difficult problem despite advances in tools.  The combination of rheology with other probes such as visualization or NMR can enable big advances.  One example would be to study shear thickening and flow in polymeric composites.  Measuring local rheological properties of a material, active microrheology is a new possibility.  It is based on using active probe particles that are sensitive to the local conditions.  Another route would be to use viscosity-sensitive dyes to detect local properties.

At the level of creating new materials, there is the need to develop rapid prototyping techniques.  The low cost and availability of 3D printers has made it possible to create particles at a small (but still not microscopic) size.  This capability opens up new types of experiments.  However, it is a slow process to manufacture particles in bulk quantities.  Thus, one advance would be to create such particles more rapidly with much smaller feature sizes.  Super-Resolution Additive Manufacturing (SRAM) would be useful to many areas of science and technology and would be a unique contribution of soft-matter science.

When manufacturing new materials, one often confronts problems associated with the fundamental properties of the raw materials.  For example, flow instabilities, rheology, and interfacial properties may limit the feature size and resolution that can be achieved.  Understanding this collective behavior is one way in which soft-matter research continues to play an important role in the development of the next stage of material fabrication.

Other areas of instrumentation development can come from exploiting developments in consumer electronics industry.  As already mentioned, GPUs, magnetic resonance imaging and X-ray tomography have entered the toolkit of soft-matter experimentalists.  One new area of cross fertilization could include the use of ultrasound probes to study density and flow behavior deep within a suspension \cite{Han2016}.

\textit{Microfluidics}:  Microfluidics is one enabling technology that deserves particular attention as a bridge between basic and applied soft-matter science \cite{Squires2005}.  It relies on powerful and cheap photolithography techniques to cast devices that control the flow of material.  Microfluidics is an elegant, effective and precise way to organize small amounts of matter and to control composition, temperature and external fields in space and time.  It is a flexible method for material synthesis and allows a large number of experiments to be performed while consuming only small amounts of raw material.  It is particularly useful for exploring length scales between $10\mu m$ and $100\mu m$.  Because these length scales are small, there is rapid transport of heat and matter and hence the ability to induce rapid changes in the material such as mixing, dissolution and phase transitions.  Microfluidics is also a powerful tool for understanding biology.  It allows the design of \textit{ex-vivo} systems to test hypotheses of the structure-function relation from the sub-cellular to the collective scale of many cells.  These are just a few of the reasons that microfluidics has become an essential and adaptable tool in soft-matter research.  Microfluidics is at the beginning of its development and many directions can be taken to produce new applications.  We will mention here only a few of them.

Current microfluidic devices are designed for benign environments. There is a need to develop devices that operate reliably in inhospitable environments ranging from human bodies in which the devices generate immune response, to the harsh conditions presented by chemical synthesis applications involving organic solvents. 

One obvious direction is to scale down the size of the devices.  Practical applications arise from the flow of fluids inside nanotubes composed of boron nitrides, zinc oxide, and carbon nanotubes.  Applications with great societal need are water purification (\textit{e.g.}, desalination, removal of heavy metals, etc.) and efficient catalysis for fuels.  However reducing the size introduces new challenges.   The large surface to volume ratios raise the fundamental question of how to make entities move when surface forces dominate.  At the nanoscale, pressure gradients are too small to drive flow.  One well developed alternative is electrical-driven transport.  Other means include using acoustic, optic, and magnetic fields.  The general question emerges of how soft matter interacts with these applied fields.  At the nanoscale, there is a need for understanding interactions at the molecular level and there needs to be a bridge from the continuum description to molecular physical chemistry.  

In order to make efficient use of these methods, students and other researchers in soft matter must get trained in the techniques.  The neutron scattering community has been effective at establishing courses for those using neutron sources.  This can serve as a guide for how to train the next generation of soft-matter scientists.  The community needs to include soft-matter topics in introductory curricula in order to make full use of the burgeoning opportunities in this area of science and technology.

\section{\label{sec:level1}Outreach\protect\\}

It is the responsibility of investigators to communicate their results to other researchers who are working in the same area.  However, this is only one aspect of scientific communication.  In addition to the research community, there is a much broader audience that should be addressed that includes students, industry, government, and the general public of interested individuals who are not necessarily expert in science (much less in a specific area).  For all of these communities, soft matter has a special role to play because it deals not only with far-reaching scientific ideas but also with topics that are accessible.  

Soft-matter is a good place to introduce students to research.  Because students can see, and sometimes even feel, the phenomenon, they can come up to speed relatively rapidly.  Soft matter encourages students to query ``why or how does this phenomenon behave the way it does?''  Moreover, the research often provides tangible output \textendash~ stunning visual images of structure or dynamics.  These characteristics have been effective at attracting diverse students into science broadly and more specifically into the field of soft matter.  The importance of the science \textendash~ to advancing applications, to addressing broad cross-cutting scientific issues, to training students and creating a diverse workforce \textendash~ should be made clear to those paying for the research.  

Because soft matter confronts issues found in our daily lives, the problems are approachable and not hidden inside a ``black box''.  The interested layperson can get a genuine opportunity to understand a science problem, understand the way the experiment is set up, and ultimately to understand the solution.  Few other areas of science can accomplish this.  In many cases, while the public can often spout some of the jargon that goes along with the science, it is not always clear that they have a real understanding of what that jargon means or how it was arrived at and how the ideas have been or could be tested.  Soft-matter research can be an antidote to such non-scientific understanding. 

It is often a comfort to hide behind obscurity and complexity; \textit{soft matter cannot take that road!} While soft-matter can be accessible to individuals with all levels of training, it is all the more essential to convey its importance.  The apparent simplicity of the systems, can lead to the impression that the research is trivial, easy, or unimportant.  One challenge is therefore to convey, correctly and without undue hype, the importance of the basic science and the importance of the applications.  It is hard to argue with a good technology.   

An effective way to accomplish this is to show the usefulness of the research to industrial colleagues by organizing short courses on different topics.  What is particularly needed is to create stronger ties with different industrial partners.  While these exist between a few individual research groups and industrial affiliates, it is less true for a majority of soft-matter groups.  It would be one important step forward to create a stronger bond between these two vital areas.

\begin{figure}
\includegraphics[width=0.85\columnwidth]{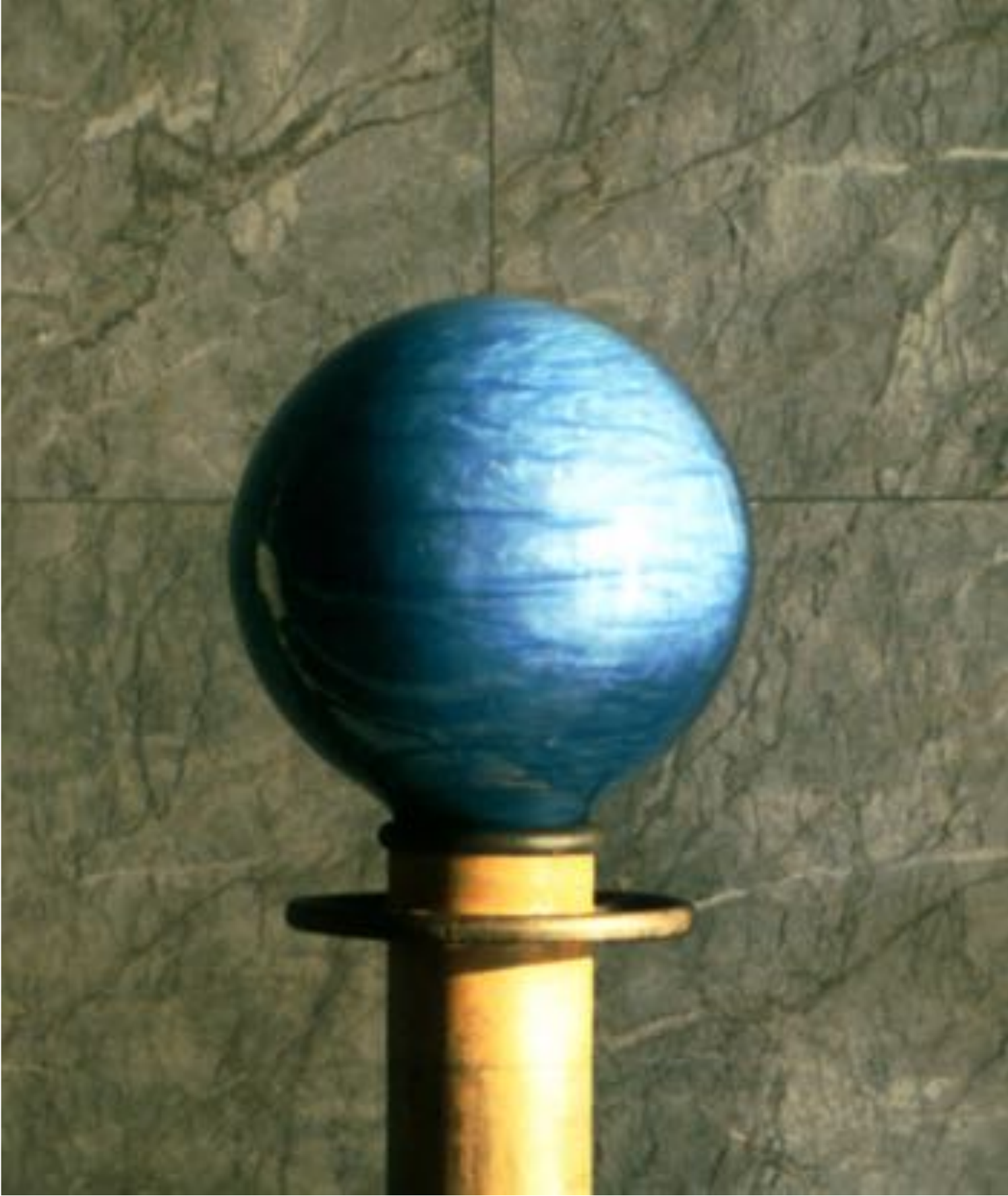}
\caption[]{\label{fig: Figure_Kahnsculpture}
Turbulent Orb by Ned Kahn displayed in Battery Park City, New York in 1990. The sculpture consists a large, rotatable, spherical glass vessel, filled with a deep-blue fluid that reveals the currents within.  From \cite{Kahn1990}.}
\end{figure}

There are a number of specific ways that soft matter can achieve the goal of explaining science to a general audience.  Some of these are a continuation of what has been so successful in the past.  It should continue to exploit the visual nature of the research with captivating imagery and video.  Because soft-matter phenomena appear on all scales, many phenomena can be displayed with hands-on demonstrations.  These can be in the form of museum exhibits where soft matter has long had a presence.  Many science museums now have excellent exhibits devoted to the intriguing behavior of soft matter.  It can be displayed in curated public spaces as was the sculpture in Fig.~\ref{fig: Figure_Kahnsculpture}.

Another area in which soft-matter research can make a difference is in working with science teachers in neighborhood schools.  Here there is the opportunity to design kits for teachers and to bring teachers into the laboratory for summer immersion in research.  Such activities can introduce students to exciting concepts in various areas, including extreme mechanics, robotics, and microfluidics.  Likewise, bringing visitors on tours of research laboratories has been a low-key but very effective way to communicate.  Grade-school students lucky enough to have had such an opportunity have sometimes advanced on to research internships early in their education. 

More recent opportunities for outreach are enabled by social-media outlets and blogs on the internet.  Fascinating websites devoted to soft-matter phenomena, for example, the site ``fuckyeahfluiddynamics'', often incorporate well-researched topics into interesting videos.  There are also special opportunities to form public networks: one idea would be to create ``big-data'' sets that could be accessed by the general public for ``crowd analysis''.

\section{\label{sec:level1}Concluding Remarks\protect\\}

Soft matter addresses issues that are broadly relevant to many areas of science.  It is, for example, arguable that understanding far-from-equilibrium behavior \textendash~a topic central to soft matter science \textendash~is the deepest and most important question facing physics today.  Soft matter has had an exciting history of discovery and is also a fertile ground for future research breakthroughs.  New tools and new ideas continue to emerge from the laboratories devoted to its study.  This report has touched on only a few of the directions where this research may lead.  Because it is broadly relevant to so many disciplines, to the way scientists can interact with the public, and to the training of a diverse next generation of students, soft-matter experiment occupies a very special position in the current scientific endeavor.  The outlook is indeed bright for continued productive and exciting activity.

\begin{acknowledgments}
The author thanks Paul Sokol of the NSF for the opportunity and encouragement to organize and write up the results of the workshop.  All the workshop attendees (listed in the Appendix) provided their wisdom about the topics covered.  I am particularly grateful to the Scientific Advisory Committee and to the session chairs for organizing and outlining the discussions.  The workshop, on which this report is based, was supported by the NSF DMR-1560936 and NSF MRSEC DMR-1420709.
\end{acknowledgments}

\appendix

\section{Participants in the workshop held in Arlington VA on January 30-31, 2016 on which this report is based}

\subsection{\label{app:subsec}Scientific Advisory Committee\protect\\}

\noindent Paul Chaikin 		(New York University)
\\Douglas Durian 	(University of Pennsylvania) 
\\Sharon Glotzer 		(University of Michigan)
\\Anette (Peko) Hosoi	(Massachusetts Inst. of Technology)
\\Heinrich Jaeger		(University of Chicago)
\\Howard Stone 		(Princeton University)

\subsection{\label{app:subsec}Attendees\protect\\} 

\noindent Robert Behringer	(Duke University)
\\Dan Blair 	(Georgetown University)
\\Paul Chaikin	(New York University) 

\textit{Session chair: Colloids}
\\Xiang Cheng	(University of Minnesota)
\\Noel Clark		(University of Colorado Ð Boulder)  

\textit{Session chair: Liquid crystals}
\\Itai Cohen		(Cornell University)  

\textit{Session chair: Mechanical meta-materials}
\\Karen Daniels	(North Carolina State University)
\\Zvonimir Dogic	(Brandeis University)  

\textit{Session chair: Active matter and biomaterials}
\\Doug Durian	(University of Pennsylvania)  

\textit{Session chair: Granular and foams}
\\Seth Fraden	(Brandeis University)  

\textit{Session chair: Microfluidics}
\\Dan Goldman	(Georgia Inst. of Technology)
\\David Grier	(New York University)
\\Anette (Peko) Hosoi	(Massachusetts Inst. of Technology)  

\textit{Session chair: Fluids}
\\William Irvine	(University of Chicago)
\\Heinrich Jaeger		(University of Chicago)  

\textit{Session chair: Outreach}
\\Ilona Kretzschmar	(City College of New York)  

\textit{Session chair: Instrumentation}
\\Dan Lathrop	(University of Maryland)
\\Wolfgang Losert	(University of Maryland) 
\\Mahesh Mahanthappa	(University of Minnesota)
\\Tom Mason	(University of California, Los Angeles)  

\textit{Session chair: Packing}
\\Gareth McKinley	(Massachusetts Inst. of Technology)
\\Sidney Nagel 	(University of Chicago)
\\Teri Odom		(Northwestern University)
\\Corey O'Hern	(Yale University)  

\textit{Session chair: Simulations and big data}
\\Chinedum Osuji	(Yale University)
\\Mark Robbins	(Johns Hopkins University)
\\Tom Russell	(University of Massachusetts Ð Amherst)  

\textit{Session chair: Polymers and rheology}
\\Todd Squires	(University of California, Santa Barbara)
\\Kathleen Stebe		(University of Pennsylvania)  

\textit{Session chair: Pattern formation}
\\Jeff Urbach	(Georgetown University)
\\Eric Weeks	(Emory University)  

\textit{Session chair: Glasses and jamming}
\\David Weitz	(Harvard University)  

\textit{Session chair: Self assembly}
\\Arjun Yodh	(University of Pennsylvania)


\bibliography{OpportunitiesBibliography.bib}

\end{document}